%% file: epimom_inference.tex
\documentclass[12pt,letterpaper]{article}

\usepackage[letterpaper,margin=1in]{geometry}
\usepackage[T1]{fontenc}
\usepackage[utf8]{inputenc}
\usepackage{lmodern}
\usepackage{amsmath}
\usepackage{amssymb}
\usepackage{amsthm}
\usepackage{graphicx}
\usepackage{xcolor}
\usepackage{hyperref}
\usepackage{xr}
\usepackage{tikz}
\usetikzlibrary{arrows.meta,positioning,shapes.geometric}
\usepackage[capitalise,nameinlink]{cleveref}

\numberwithin{equation}{section}
\newcommand{\hideshowlabels}{}
\input{shared/epimom_preamble}
\epimomarxivtrue
\input{shared/arxiv_external_labels_theory}
\input{epimom_inference/epimom_inference_metadata}

\input{shared/epimom_article_metadata}
\let\hypercref\labelcref

\renewcommand{\beginsection}[2]{\section{#1}\label{#2}}
\renewcommand{\beginsubsection}[2]{\subsection{#1}\label{#2}}
\renewcommand{\beginsubsubsection}[2]{\subsubsection{#1}\label{#2}}
\renewcommand{\beginappendix}[2]{\section{#1}\label{#2}}

\epimomarticlemetadata

\begin{document}

\maketitle

\begin{abstract}
  \input{epimom_inference/epimom_inference_abstract}
\end{abstract}

\epimomarticlekeywords


\input{epimom_inference/epimom_inference_main}

\clearpage
\section*{Declaration of AI use}
\msaiuse

\ifanonymous\else
  \section*{Funding}
  \msfunding

  \section*{Acknowledgements}
  \msacknowledgements
\fi

\bigbreak\bigbreak\bigbreak
\appendix
\section*{APPENDICES}
\numberwithin{equation}{section}
\renewcommand{\theequation}{\thesection\arabic{equation}}
\input{epimom_inference/epimom_inference_appendices}

\bibliographystyle{RS}
\bibliography{epimom-refs}

\end{document}

%% file: shared/epimom_preamble.tex
\usepackage{framed}    
\usepackage{xspace}    
\usepackage{multicol}  

\newif\ifepimomarxiv
\epimomarxivfalse

\input{shared/epimom_build_modes}

\iflinenumbers
  \usepackage{lineno}
  
  \makeatletter
  \@ifclassloaded{rsproca_new}{%
    \AddToHook{cmd/maketitle/after}[epimom/linenumbers]{\linenumbers}%
  }{%
    \@ifclassloaded{rspb_current_style}{%
      \def\epimom@makelinenumber{%
        \ifnum\value{page}=\@ne
          \makeLineNumberRight
        \else
          \makeLineNumberLeft
        \fi
      }%
      \let\makeLineNumberRunning\epimom@makelinenumber
      \AddToHook{cmd/maketitle/after}[epimom/linenumbers]{\linenumbers}%
    }{%
      \AtBeginDocument{\linenumbers}%
    }%
  }%
  \makeatother
\fi

\ifdefined\hideshowlabels
\else
  \definecolor{labelgray}{gray}{0.8}
  
  \usepackage{showlabels}
  
\fi

\newcommand{\ie}{\textit{i.e., }}
\newcommand{\eg}{\textit{e.g., }}

\newcommand{\RE}{renewal equation\xspace}

\newcommand{\mathscale}[2]{%
  \mathchoice
    {\scalebox{#1}{$\displaystyle #2$}}%
    {\scalebox{#1}{$\textstyle #2$}}%
    {\scalebox{#1}{$\scriptstyle #2$}}%
    {\scalebox{#1}{$\scriptscriptstyle #2$}}%
}

\newcommand{\R}{{\mathcal R}}
\newcommand{\Rn}{\R_{\mathscale{\mymathscale}{0}}}
\newcommand{\Reff}{\R_{\mathscale{\mymathscale}{{\mathrm{eff}}}}}
\newcommand{\Ra}{\R_{\mathscale{\mymathscale}{\alpha}}}

\newcommand{\FoI}{F}        
\newcommand{\aoi}{\alpha}   
\newcommand{\aoidum}{\aoi'}

\newcommand{\inc}{\iota}    
\newcommand{\cuminc}{\underline{\inc}} 

\newcommand{\Volterra}{V}
\newcommand{\CFPvar}{\texttt{\upshape CFP}}
\newcommand{\gimean}{\mu}
\newcommand{\gisd}{\sigma}
\newcommand{\Erate}{\gamma_{\!\scalebox{0.5}{$E$}}}
\newcommand{\Irate}{\gamma_{\!\scalebox{0.5}{$I$}}}
\newcommand{\Eprop}{Y_{\!\scalebox{0.5}{$E$}}}
\newcommand{\Iprop}{Y_{\!\scalebox{0.5}{$I$}}}

\newcommand{\Tlat}{T_{\rm lat}}
\newcommand{\Tinf}{T_{\rm inf}}
\newcommand{\Winv}{\mathscr{E}} 
\newcommand{\Wp}{W_{\!\smallplus}}
\newcommand{\Wm}{W_{\!\smallminus}}
\newcommand{\Wpm}{W_{\!\smallplusminus}}

\newcommand{\mymathscale}{0.7} 

\newcommand{\tauinit}{\tau_{\rm i}}
\newcommand{\xinit}{x_{\rm i}}
\newcommand{\yinit}{y_{\rm i}}

\newcommand{\epropinit}{y^{\scalebox{0.5}{$E$}}_{\mathrm{i}}}
\newcommand{\ipropinit}{y^{\scalebox{0.5}{$I$}}_{\mathrm{i}}}
\newcommand{\incinit}{\inc_{\rm i}}

\newcommand{\taupeak}{\hat{\tau}}
\newcommand{\xpeak}{\hat{x}}
\newcommand{\ypeak}{\hat{y}}

\newcommand{\tauincmax}{\hat{\tau}_{\inc}} 

\newcommand{\smallplus}{{\mathscale{\mymathscale}{+}}}
\newcommand{\smallminus}{{\mathscale{\mymathscale}{-}}}
\newcommand{\smallplusminus}{{\mathscale{\mymathscale}{\pm}}}

\newcommand{\xp}{x^{\smallplus}}  
\newcommand{\xm}{x^{\smallminus}} 
\newcommand{\xpm}{x^\smallplusminus}

\newcommand{\zp}{z^{\smallplus}} 
\newcommand{\zptot}{\zp_{\mathscale{\mymathscale}{\mathrm{tot}}}}
\newcommand{\zm}{z^{\smallminus}} 
\newcommand{\lamp}{\lambda^{\!{\smallplus}}}
\newcommand{\lamm}{\lambda^{\!{\smallminus}}}
\newcommand{\lampm}{\lambda^{\!{\smallplusminus}}}

\newcommand{\tspanel}[1]{{\bfseries\textsf{\MakeLowercase{#1}}}\xspace}
\newcommand{\pppanel}{\tspanel{C}} 

\newcommand{\dee}{{\rm d}}
\newcommand{\dd}[2]{{\frac{\dee{#1}}{\dee{#2}}}}

\newcommand{\ddx}[1]{\dd{#1}{x}}
\newcommand{\ddt}[1]{\dd{#1}{t}}
\newcommand{\ddtau}[1]{\dd{#1}{\tau}}

\usepackage{mathrsfs} 
\DeclareFontFamily{U}{rsfs}{\skewchar\font127}
\DeclareFontShape{U}{rsfs}{m}{n}{%
   <-6> rsfs5
   <6-8> rsfs7
   <8-> rsfs10
}{}
\newcommand{\Lapsym}{{\mathscr{L}}}
\newcommand{\Lap}[1]{{\Lapsym}\!\left[#1\right]}

\newcommand{\Lappm}{\Lapsym_{\!\smallplusminus}}
\newcommand{\Lapm}{\Lapsym_{\!\smallminus}}

\newcommand{\term}[1]{{\bfseries\slshape #1}}

\makeatletter

\AddToHook{cmd/appendix/after}[epimom/cleveref-appendix-types]{%
  \crefalias{section}{appendix}%
  \crefalias{subsection}{subappendix}%
  \crefalias{subsubsection}{subsubappendix}%
}

\newcommand{\beginsection}[2]{%
  \section*{#1}%
  \phantomsection%
  \def\@currentlabelname{#1}%
  \label{#2}%
  \pdfbookmark[1]{#1}{#2}%
}
\newcommand{\beginappendix}[2]{\beginsection{#1}{#2}}

\newcommand{\beginsubsection}[2]{%
  \subsection*{#1}%
  \phantomsection%
  \def\@currentlabelname{#1}%
  \label{#2}%
  \pdfbookmark[2]{#1}{#2}%
}

\newcommand{\beginsubsubsection}[2]{%
  \subsubsection*{#1}%
  \phantomsection%
  \def\@currentlabelname{#1}%
  \label{#2}%
  \pdfbookmark[2]{#1}{#2}%
}

\makeatother

\newif\ifshowappbackrefs
\showappbackrefsfalse  

\newcommand{\mainref}[1]{\cref{#1}}

\newcommand{\theoryref}[1]{%
  \citeref[\cref*{theory:#1}]{EarnPars25_epimom_theory}%
}

\DeclareRobustCommand{\appref}[2][]{%
  \if\relax\detokenize{#1}\relax
    \cref{#2}%
  \else
    \hypertarget{appref:#1}{}\cref{#2}%
  \fi
}

\newcommand{\appbackref}[2]{%
  \hyperlink{appref:#1}{#2}%
}

\newcommand{\appbackrefs}[1]{%
  \ifshowappbackrefs
    \par\smallskip
    \noindent{\color{red}{\bfseries Referenced from: }#1.}%
    \par\smallskip
  \fi
}

\newcommand{\appfilename}[1]{%
  \ifshowappbackrefs
    \par\smallskip
    \noindent{\footnotesize\textit{Appendix source file: }\texttt{\detokenize{#1}}.}%
    \par\smallskip
  \fi
}

\makeatletter

\newcommand{\mmsectionlabel}[1]{%
  \phantomsection%
  \def\@currentlabelname{Materials and Methods}%
  \label{#1}%
  \pdfbookmark[1]{Materials and Methods}{#1}%
}

\newcommand{\mmsubsection}[2]{%
  \subsection*{#1}%
  \phantomsection%
  \def\@currentlabelname{#1}%
  \label{#2}%
  \pdfbookmark[2]{#1}{#2}%
}

\newcommand{\mmsubsubsection}[2]{%
  \subsubsection*{#1}%
  \phantomsection%
  \def\@currentlabelname{#1}%
  \label{#2}%
  \pdfbookmark[2]{#1}{#2}%
}

\renewcommand{\mmsectionlabel}[1]{%
  \phantomsection%
  \def\@currentlabelname{Appendix A}%
  \label{#1}%
  \pdfbookmark[1]{Appendix A}{#1}%
}

\renewcommand{\mmsubsection}[2]{%
  \subsection{#1}%
  \phantomsection%
  \def\@currentlabelname{#1}%
  \label{#2}%
  \pdfbookmark[2]{#1}{#2}%
}

\renewcommand{\mmsubsubsection}[2]{%
  \subsubsection{#1}%
  \phantomsection%
  \def\@currentlabelname{#1}%
  \label{#2}%
  \pdfbookmark[2]{#1}{#2}%
}

\makeatother

\let\citeref\cite

\newcommand{\nocomment}[3]{}

\usepackage[normalem]{ulem}
\newcommand{\stkout}[1]{{\color{red}\ifmmode\text{\sout{\ensuremath{#1}}}\else\sout{#1}\fi}}

\newcommand{\Rnmin}{\R_{\mathscale{\mymathscale}{0,\textrm{\upshape min}}}}

\newcommand{\Rnstar}{{\Rn^\ast}}

\usepackage{mathtools} 

\newcommand{\zplim}{\zp_{\mathscale{\mymathscale}{\mathrm{lim}}}}
\newcommand{\Rnstarmin}{\R^*_{\mathscale{\mymathscale}{\mathrm{0},\vee}}}
\newcommand{\Rnstarmax}{\R^*_{\mathscale{\mymathscale}{\mathrm{0},\wedge}}}

\usepackage{amssymb} 

%% file: shared/epimom_build_modes.tex
\makeatletter
\@ifundefined{ifanonymous}{%
  \newif\ifanonymous
  \anonymousfalse
}{}
\@ifundefined{iflinenumbers}{%
  \newif\iflinenumbers
  \linenumbersfalse
}{}
\makeatother

%% file: epimom_inference/epimom_inference_metadata.tex
\newcommand{\mstitle}{Jointly estimating transmissibility and prior immunity from epidemic time series}

\newcommand{\msauthorone}{David J.\,D.\ Earn}
\newcommand{\msauthortwo}{Todd L.\ Parsons}

\newcommand{\msarticleauthors}{%
  \msauthorone\thanks{Department of Mathematics \& Statistics and
  M.\,G.\ DeGroote Institute for Infectious Disease Research, McMaster
  University, 1280 Main Street West, Hamilton, Ontario, L8S 4K1, Canada.
  Email: \href{mailto:\mscorrespondingemail}{\mscorrespondingemail}.
  ORCID: \href{https://orcid.org/0000-0002-7562-1341}{0000-0002-7562-1341}.}
  \and
  \msauthortwo\thanks{LPSM, Sorbonne Universit\'e, CNRS UMR 8001,
  Paris 75005, France. ORCID:
  \href{https://orcid.org/0000-0002-2599-8415}{0000-0002-2599-8415}.}%
}

\newcommand{\mscorrespondingemail}{earn@math.mcmaster.ca}

\newcommand{\mskeywords}{epidemic momentum, prior immunity, epidemic inference, basic reproduction number, 1918 influenza pandemic}

\newcommand{\msfunding}{We were supported by the Fields Institute for Research in
  Mathematical Sciences; the contents of this paper are solely the
  responsibility of the authors and do not necessarily represent the
  official views of the Institute.  DJDE was supported by an NSERC
  Discovery Grant.}
\newcommand{\msacknowledgements}{We are grateful to Ben Bolker, Caroline Colijn, Jonathan Dushoff,
  Maya Earn, Mark Lewis, Junling Ma, David Price, and Steve Walker for
  comments and discussions.  Deterministic and stochastic simulations,
  and deconvolution of mortality to incidence, were facilitated by
  convenient functions in the \textsf{fastbeta} R package, written by
  Mikael Jagan.}

\newcommand{\msaiuse}{We used ChatGPT and OpenAI Codex to assist with
  R coding, \LaTeX\ coding, and language editing.  All scientific
  arguments, mathematical derivations, analyses, interpretations, and
  conclusions were developed by the authors, who take full
  responsibility for the final manuscript.}

%% file: shared/epimom_article_metadata.tex
\newcommand{\epimomarticlemetadata}{%
  \title{\mstitle}%
  \ifanonymous
    \author{}%
    \hypersetup{%
      pdfauthor={},
      pdftitle={\mstitle},
      pdfsubject={Anonymous review manuscript},
      pdfkeywords={\mskeywords}}%
  \else
    \author{\msarticleauthors}%
    \hypersetup{%
      pdftitle={\mstitle},
      pdfsubject={},
      pdfkeywords={\mskeywords}}%
  \fi
  \date{}%
}

\newcommand{\epimomarticlekeywords}{%
  \par\medskip
  \noindent\textbf{Keywords:} \mskeywords\par
}

%% file: epimom_inference/epimom_inference_abstract.tex
Infectious disease time series are often used to estimate a pathogen's basic reproduction number, $\Rn$.  However, fits of epidemic models to time series conflate pathogen transmissibility with pre-existing population immunity, so only the \emph{effective} reproduction number, $\Reff$, can be inferred.  This composite parameter is the product of the underlying $\Rn$ and the pre-epidemic susceptible fraction, $\xm$. We show that a conservation law associated with \emph{epidemic momentum}---prevalence weighted by potential to infect---makes it possible to disentangle transmissibility from prior immunity and to infer $\Rn$ and $\xm$ separately from a single epidemic time series. We test the methodology using stochastic epidemic simulations, and illustrate the approach with a reappraisal of influenza transmissibility during the 1918 pandemic, estimating rather than assuming the degree of prior population immunity.  For the autumn wave in Philadelphia, USA, we find $\Rn\approx2.7$ and $\xm\approx0.8$, implying that about 20\% of the population was already immune before that wave, plausibly as a result of infection during the spring 1918 herald wave.

%% file: epimom_inference/epimom_inference_main.tex
\providecommand{\epimomfirstpagebreak}{}

\beginsection{Introduction}{sec:intro}

Epidemic time series are commonly used to infer how well an infection
spreads, as measured by the basic reproduction number, $\Rn$---the
expected number of secondary infections caused by a typical infected
individual in a fully susceptible population.  A standard approach
estimates the initial exponential growth rate and combines it with an
estimated or assumed distribution of intrinsic generation
intervals---the delays between infection of an individual and the
secondary infections caused by that individual---to obtain a
reproduction number \citeref{WallLips07,Ma+14,Ma20}.  More precisely,
if $\xm$ is the fraction of the population that was susceptible before
the epidemic began, initial growth identifies the effective
reproduction number $\Reff=\Rn\xm$, rather than $\Rn$
\citeref{McCaw+2009}.  Early growth alone does not separate pathogen
transmissibility from pre-existing population immunity, because
different combinations of $\Rn$ and $\xm$ can give the same $\Reff$.

The fraction of the population that was already immune when the
epidemic began is $\zm=1-\xm$.  Allowing for pre-existing immunity
($\xm<1$) changes both the transmissibility inferred for the pathogen
and the estimated proportion infected during the focal epidemic,
$\zp=\xm-\xp$, where $\xp$ is the fraction susceptible after the
epidemic.  Thus, assuming that the population was initially fully
susceptible confounds the interpretation of transmissibility, prior
immunity, and final size.

Epidemic momentum \cite{EarnPars25_epimom_theory} provides the
additional information needed to separate $\Rn$ from
$\xm$. Conceptually, epidemic momentum measures the infectious
potential carried by the population, weighting past infections by
their remaining potential to transmit. It can be constructed from an
incidence time series and the intrinsic generation-interval
distribution $g(\aoi)$, where $\aoi$ is the time since infection. The
susceptible fraction and epidemic momentum satisfy a conservation
law \theoryref{sec:FI}, which supplies the additional relation needed
to determine $\Rn$ and $\xm$ separately.

\epimomfirstpagebreak

In this paper, we exploit epidemic momentum to infer $\Rn$ and $\xm$
separately from the same epidemic time series.  We assess this
momentum-based method for inferring transmissibility and prior
immunity using stochastic epidemic simulations for which both
quantities are known, and then apply it to the autumn wave of the 1918
influenza pandemic in Philadelphia. The application illustrates the
biological consequences of estimating, rather than assuming, the
population immunity present before an epidemic.

The inference proceeds from an observed incidence time series, or from
a time series from which incidence can be reconstructed (\eg
mortality \cite{Gold+09}), together with an assumed or estimated
intrinsic generation interval distribution $g(\aoi)$.  As in standard
growth-rate inference \cite{WallLips07}, we combine $g(\aoi)$ with the
estimated initial growth rate to infer the product $\Reff=\Rn\xm$.
The incidence history and $g(\aoi)$ then determine epidemic momentum
and the effective reproduction number through time. Evaluating the
conservation law using these quantities yields the additional relation
needed to separate $\Rn$ from $\xm$; prior immunity and final size
follow from the inferred susceptible fractions.

We first introduce the renewal-equation formulation of epidemic
models, from which we derive the relations underlying the
momentum-based method. We then assess its accuracy using stochastic simulations,
apply it to main-wave data recorded in Philadelphia in 1918, and
discuss the biological interpretation and limitations of the
inferences.

\beginsection{Methods}{sec:methods}

\beginsubsection{Kermack and McKendrick's renewal equation}{sec:re}

We formulate the momentum-based method using the \RE, which represents
a broad class of epidemic models without requiring a particular
compartmental structure \citeref{KermMcKe27,Bred+12,Cham+18}. This
formulation is particularly convenient for inference because it is
written in terms of incidence---the quantity observed, or
reconstructed, from an epidemic time series---and the intrinsic
generation-interval distribution, which must already be estimated or
assumed to convert an epidemic growth rate into a reproduction
number \cite{WallLips07}.

We measure time in units of a convenient epidemiological timescale,
often the mean infectious period, so that $\tau$ is dimensionless and
the equations can be written in terms of $\Rn$ rather than separate
transmission and recovery-rate parameters.  We write $X(\tau)$ for the
susceptible fraction at time $\tau$, and $\inc(\tau)$ for the
incidence, the fraction of the population newly infected per unit
time.  The probability density $g(\aoi)$ specifies the distribution of
the \term{intrinsic generation interval}, the delay between infection
of an individual and a secondary infection caused by that
individual \citeref{Sven07,ChamDush15}.  In dimensionless time, the
\term{\RE} is
\begin{subequations}\label{eq:re}
\begin{align}
  \ddtau{X}
  &\;=\;
  -\inc(\tau),
  \label{eq:re;X}
  \\
  \inc(\tau)
  &\;=\;
  \Rn X(\tau)
  \int_{-\infty}^{\tau}
    \inc(s)g(\tau-s)\,\dee s.
  \label{eq:re;inc}
\end{align}
\end{subequations}
The first equation identifies incidence with the rate at which
susceptibles are depleted.  In the second equation, infections that
occurred at time $s$ contribute to incidence at time $\tau$ according
to the generation-interval density $g$ evaluated at their infection
age, $\tau-s$.  Since incidence and $g$ are non-negative, the integral in
\cref{eq:re;inc} is positive whenever $\inc(s)$ has been positive for
more than an isolated instant at times when $g(\tau-s)>0$.  The
effective reproduction number at time $\tau$ can therefore be written
\begin{equation}\label{eq:re.Reff}
  \Reff(\tau)
  \;=\;
  \Rn X(\tau)
  \;=\;
  \frac{\inc(\tau)}{
    \displaystyle\int_{-\infty}^{\tau}
      \inc(s)g(\tau-s)\,\dee s} \,.
\end{equation}
Multiplying incidence by a constant introduces the same factor in the
numerator and denominator, so $\Reff(\tau)$ does not depend on the
overall scale of the incidence curve.

In the framework represented by \cref{eq:re}, differences in
infection-age structure among epidemic models are captured by
$g(\aoi)$.  Standard examples of generation-interval distributions,
including those corresponding to the SIR and SEIR models, are given
in \cref{tab:models}.  Other choices of $g(\aoi)$ represent
models with more general infection-age structure \cite{Cham+18}.  A
fuller account of the \RE, its initial history, and its relation to
compartmental models is given in \theoryref{sec:re}.

The lower limit of integration in \cref{eq:re;inc} makes explicit that
incidence at time $\tau$ depends on the preceding incidence history.
This history enters the inference in two ways.  First, exponentially
growing incidence in the early epidemic tail yields the Euler--Lotka
relation that identifies $\Reff=\Rn\xm$.  Second, the incidence
history and $g(\aoi)$ determine epidemic momentum, which weights past
infections by their remaining reproductive potential. In practice,
observations begin at a finite time, but the early exponential
approximation also allows the unobserved part of the incidence history
to be included in the momentum calculation.  We begin with the
relationship between the asymptotic exponential growth and decay rates
and transmissibility.

\beginsubsection{Tail exponents and reproduction numbers}{sec:WL}

In the initial ($\tau\to-\infty$) and final
($\tau\to+\infty$) phases of an epidemic, incidence grows and decays
exponentially, 
\begin{equation}
  \inc(\tau)
  \;\propto\;
  e^{\lampm\tau},
  \label{eq:inc.exp}
\end{equation}
where $\lamm>0$ applies near the start of the epidemic and $\lamp<0$
applies near the end; we refer to $\lampm$ as the \term{tail
exponents}.  During these early and late phases, incidence is small
and the susceptible fraction is nearly constant, at $\xm$ and $\xp$,
respectively.  Substituting \cref{eq:inc.exp} into \cref{eq:re;inc},
rearranging, and cancelling common factors gives the Euler--Lotka
equation
\citeref{DublLotk25,Fell41,KeyfCasw05},
\begin{equation}
  \frac{1}{\Rn\xpm}
  \;=\;
  \int_0^\infty
    e^{-\lampm\aoi}g(\aoi)\,\dee\aoi
  \;\equiv\;
  \Lap{g}(\lampm)
  \;\equiv\;
  \Lappm.
  \label{eq:lamLaplace}
\end{equation}
\Cref{eq:lamLaplace} therefore identifies the effective reproduction
number in the initial exponential-growth limit,
\begin{equation}\label{eq:Reff.initial}
  \Reff(-\infty)
  \;=\;
  \Rn\xm
  \;=\;
  \frac{1}{\Lapm}.
\end{equation}
Setting $\xm=1$ yields the formula of Wallinga and
Lipsitch \citeref[Equation~(2.7)]{WallLips07}, which is often used with
empirical estimates of $\lamm$ to infer $\Rn$
\citeref{WallLips07,Ma+14,Ma20}.  If $\xm<1$, the same calculation
instead estimates $\Reff(-\infty)=\Rn\xm$ \citeref{McCaw+2009}.

Estimating $\xm$ empirically is sometimes possible
(\eg \citeref{Ston+07}).  Computationally demanding and/or
model-specific methods that attempt to infer or constrain $\xm$
indirectly from the observed epidemic data have also been proposed
(see \cref{sec:discuss.prior}).  Our approach obtains the additional
information directly from epidemic momentum, which we now define
precisely.

\beginsubsection{Constructing epidemic momentum from incidence}{sec:epimom}

At infection age $\aoi$, the upper tail of the generation-interval
distribution is the fraction of an individual's intrinsic transmission
potential that remains.  Multiplying this fraction by $\Rn$, we define
the \term{reduced reproduction number}
\begin{equation}
  \Ra
  \;=\;
  \Rn
  \int_{\aoi}^{\infty}
    g(\aoidum)\,\dee\aoidum .
  \label{eq:Ra}
\end{equation}
Thus $\Ra$ is the expected number of future secondary infections caused
by an individual of infection age $\aoi$ in a fully susceptible
population, while $\Ra/\Rn$ is the fraction of that individual's
reproductive potential that remains (which depends only on $g(\aoi)$,
not $\Rn$).

At time $\tau$, the cohort infected at time $\tau-\aoi$ contributes
$\inc(\tau-\aoi)\,\dee\aoi$ to the infection-age distribution.
Weighting each cohort by its remaining reproductive potential, we
obtain the \term{epidemic momentum} \cite{EarnPars25_epimom_theory},
\begin{equation}
  Y(\tau)
  \;=\;
  \int_0^\infty
    \inc(\tau-\aoi)
    \frac{\Ra}{\Rn}\,\dee\aoi .
  \label{eq:re.Y}
\end{equation}
For the SIR and SEIR models, epidemic momentum coincides with the
total prevalence of infection.  
More generally, epidemic momentum is a weighted prevalence: each
currently infected individual contributes according to their remaining
potential to infect, so individuals at different infection ages carry
different weights.

An equivalent representation is useful when epidemic observations are
expressed as cumulative incidence.  Writing
$\cuminc(\tau)=\int_{-\infty}^{\tau}\inc(s)\,\dee s$ and integrating
\cref{eq:re.Y} by parts, using
$\dd{}{\aoi}(\Ra/\Rn)=-g(\aoi)$, gives
\begin{equation}
  Y(\tau)
  \;=\;
  \cuminc(\tau)
  -
  \int_0^\infty
    \cuminc(\tau-\aoi)g(\aoi)\,\dee\aoi .
  \label{eq:re.Y.g}
\end{equation}
Thus epidemic momentum can be computed either directly from incidence
using \cref{eq:re.Y}, or from the corresponding cumulative-incidence
curve using \cref{eq:re.Y.g}.

Both representations formally involve the complete incidence history.
In practice, suppose observations begin at time $\tauinit$ during the
initial exponential-growth phase, with
$\inc(\tauinit)=\incinit$.  Incidence before $\tauinit$ can then be
approximated by
$\inc(s)\approx\incinit e^{\lamm(s-\tauinit)}$.  Splitting
\cref{eq:re.Y} into the contributions before and after $\tauinit$ gives,
for $\tau\geq\tauinit$,
\ifdefined\epimomnarrowcolumns
  \begin{align}
    Y(\tau)
    &\;\approx\;
    \incinit e^{\lamm(\tau-\tauinit)}
    \int_{\tau-\tauinit}^{\infty}
      e^{-\lamm\aoi}
      \frac{\Ra}{\Rn}\,\dee\aoi
    \notag\\
    &\qquad
    +
    \int_0^{\tau-\tauinit}
      \inc(\tau-\aoi)
      \frac{\Ra}{\Rn}\,\dee\aoi \,.
    \label{eq:re.Y.finite}
  \end{align}
\else
  \begin{align}
    Y(\tau)
    \quad\approx\quad
    \incinit e^{\lamm(\tau-\tauinit)}
    \int_{\tau-\tauinit}^{\infty}
      e^{-\lamm\aoi}
      \frac{\Ra}{\Rn}\,\dee\aoi
    \quad
    +
    \quad
    \int_0^{\tau-\tauinit}
      \inc(\tau-\aoi)
      \frac{\Ra}{\Rn}\,\dee\aoi \,.
    \label{eq:re.Y.finite}
  \end{align}
\fi
The first term in \cref{eq:re.Y.finite} reconstructs the contribution
from infections that occurred before observations began, while the
second is computed directly from the observed incidence history.  A
Laplace-transform representation of the first term, including the
explicit SEIR expressions used in our calculations, is derived in
\appref[finite-history-momentum]{app:yinit.from.incinit.finite.int.rep}.

Consequently, the observed incidence curve, its initial growth rate
$\lamm$, and the generation-interval distribution $g(\aoi)$ determine
the epidemic momentum throughout the observed epidemic.  
Using $Y(\tau)$ from \cref{eq:re.Y.finite} together with $\Reff(\tau)$
from \cref{eq:re.Reff}, we can infer $\Rn$ and $\xm$ separately.

\beginsubsection
    {\texorpdfstring{Inferring $\Rn$ and prior population immunity using epidemic momentum}{Disentangling R0 from prior population immunity}}{sec:est.Rn.xm}

The preceding subsections provide the quantities needed to separate
$\Rn$ from $\xm$.  The rising tail exponent $\lamm$ and the intrinsic
generation-interval distribution $g(\aoi)$ determine the product
$\Rn\xm$ (\ie $\Reff$ in the limit $\tau\to-\infty$)
through \cref{eq:Reff.initial}.  At any later time $\tau$, the same
incidence history and $g(\aoi)$ determine the time evolution of
$\Reff$ through \cref{eq:re.Reff} and of the epidemic momentum $Y$
through the finite-history expression \labelcref{eq:re.Y.finite},
derived from \cref{eq:re.Y}.

Differentiating the cumulative-incidence representation
\cref{eq:re.Y.g}, using $\ddtau{\cuminc}=\inc$, gives
\begin{subequations}\label{eq:state.dynamics}
\begin{align}
  \ddtau{Y}
  &\;=\;
  \inc(\tau)
  -
  \int_0^\infty
    \inc(\tau-\aoi)g(\aoi)\,\dee\aoi \,.
  \label{eq:state.dynamics;Y.aoi}
  \\
\intertext{Setting $s=\tau-\aoi$ gives}
  \ddtau{Y}
  &\;=\;
  \inc(\tau)
  -
  \int_{-\infty}^{\tau}
    \inc(s)g(\tau-s)\,\dee s \,.
  \label{eq:state.dynamics;Y.inc}
  \\
\intertext{\Cref{eq:re.Reff} then gives}
  \ddtau{Y}
  &\;=\;
  \inc(\tau)
  \left(1-\frac{1}{\Reff(\tau)}\right) \,.
  \label{eq:state.dynamics;Y.Reff}
  \\
\intertext{Defining $\FoI(\tau)=\inc(\tau)/X(\tau)$ and substituting
$\Reff(\tau)=\Rn X(\tau)$ gives}
  \ddtau{Y}
  &\;=\;
  \left(X(\tau)-\frac{1}{\Rn}\right)\FoI(\tau) \,.
  \label{eq:state.dynamics;Y}
  \\
\intertext{\Cref{eq:re;X} and the definition of $\FoI$ give}
  \ddtau{X}
  &\;=\;
  -\inc(\tau)
  \;=\;
  -X(\tau)\FoI(\tau) \,.
  \label{eq:state.dynamics;X}
  \\
\intertext{Where incidence is positive, dividing
\cref{eq:state.dynamics;Y} by \cref{eq:state.dynamics;X} gives}
  \ddx{Y}
  &\;=\;
  -1+\frac{1}{\Rn x} \,.
  \label{eq:state.dynamics;YX}
\end{align}
\end{subequations}
Integrating \cref{eq:state.dynamics;YX} gives the first integral
\begin{equation}\label{eq:FI}
  C(x,y)
  \;=\;
  y
  +
  \xpeak\,
  \Volterra\!\left(\frac{x}{\xpeak}\right),
\end{equation}
where $\Volterra(u)=u-1-\ln u$ is the ``Volterra function''
\citeref{EarnMcCl25} and $\xpeak=\frac{1}{\Rn}$. 
Conservation of \cref{eq:FI} along epidemic trajectories is the
epidemic-momentum conservation law \cite{EarnPars25_epimom_theory}.
Evaluating \cref{eq:FI} at the pre-epidemic endpoint $(\xm,0)$ and an
arbitrary reconstructed state yields the identity
$C\bigl(X(\tau),Y(\tau)\bigr)=C(\xm,0)$, which can be written
\begin{subequations}\label{eq:CQ}
\begin{align}
  \Rn Y(\tau)
  &\;=\;
  \Volterra(\Rn\xm)
  -
  \Volterra\bigl(\Rn X(\tau)\bigr),
  \label{eq:CQ;RnX}
  \\
  &\;=\;
  \Volterra\bigl(\Reff(-\infty)\bigr)
  -
  \Volterra\bigl(\Reff(\tau)\bigr).
  \label{eq:CQ;Reff}
\end{align}
\end{subequations}
Consequently, any reconstructed state with $Y(\tau)>0$ gives
\begin{subequations}\label{eq:R.C.state}
\begin{align}
  \Rn
  &\;=\;
  \frac{
    \Volterra\bigl(\Reff(-\infty)\bigr)
    -
    \Volterra\bigl(\Reff(\tau)\bigr)}
  {Y(\tau)},
  \label{eq:R.C.state;Rn}
  \\
  \xm
  &\;=\;
  \frac{\Reff(-\infty)}{\Rn}.
  \label{eq:R.C.state;xm}
\end{align}
\end{subequations}
This identification assumes that transmissibility is constant from the
initial growth phase through time $\tau$, and that the intrinsic
generation-interval distribution $g(\aoi)$ is known.  Evaluating
\cref{eq:re.Reff,eq:re.Y} requires the incidence history preceding
$\tau$.  The portion before observations began must be reconstructed
using the initial exponential-growth approximation, as in the
finite-history expression for momentum in \cref{eq:re.Y.finite}.
Infection incidence must be calibrated on an absolute per-capita scale
to reconstruct $Y(\tau)$, whereas no such calibration is needed to
determine $\Reff(\tau)$ because the ratio in \cref{eq:re.Reff} is
invariant to an overall incidence-scale factor.

At the momentum peak $\taupeak$, positive incidence and
\cref{eq:state.dynamics;Y} give $\Reff(\taupeak)=1$, while
$Y(\taupeak)=\ypeak$ and $\Volterra(1)=0$.  Thus
\cref{eq:R.C.state} has the immediate corollary
\begin{equation}\label{eq:R.C.lampm}
  \Rn
  \;=\;
  \frac{1}{\ypeak} \,
  \Volterra\!\Big(\frac{1}{\Lapm}\Big) \,.
\end{equation}
When the momentum peak has been observed, the peak-specific
formula \labelcref{eq:R.C.lampm} has several advantages
over \cref{eq:R.C.state;Rn}.  The peak condition fixes
$\Reff(\taupeak)=1$, eliminating the need to estimate
$\Reff(\taupeak)$, while both $Y(\tau)$ and the numerator of
\cref{eq:R.C.state;Rn} are maximal at $\taupeak$
[\appref{app:peak.properties}], avoiding numerical sensitivity associated
with division by a small quantity.  Once the peak has been reliably
reconstructed, evaluating the general
formula \labelcref{eq:R.C.state;Rn} at a later state offers no
corresponding advantage.  The simulations and Philadelphia analysis
below use the peak formula \labelcref{eq:R.C.lampm}.

Once $\Rn$ has been inferred, \cref{eq:Reff.initial} gives
\begin{equation}
  \xm
  \,=\;
  \frac{1}{\Rn\Lapm}
  \quad\text{and}\quad
  \zm
  \,=\;
  1-\xm
  \;=\;
  1-\frac{1}{\Rn\Lapm}.
  \label{eq:zm.compute}
\end{equation}
Hence the estimated initial growth rate $\lamm$, the reconstructed
values of $Y(\tau)$ and $\Reff(\tau)$, and $g(\aoi)$ determine both
pathogen transmissibility ($\Rn$) and prior population immunity
($\zm$).

Classical final-size relations express the asymptotic post-epidemic
susceptible fraction $\xp$ in terms of an epidemic state specified at
a finite initial time \cite{KermMcKe27,MaEarn06,Mill2012}.  Here,
after both $\Rn$ and $\xm$ have been inferred, applying the
conservation law \labelcref{eq:FI} to the asymptotic pre- and
post-epidemic endpoints gives
\begin{equation}\label{eq:V=V}
  \Volterra(\Rn\xm)
  \;=\;
  \Volterra(\Rn\xp),
\end{equation}
or equivalently
\begin{equation}\label{eq:final.size.relation}
  \ln\!\left(\frac{\xm}{\xp}\right)
  \;=\;
  \Rn(\xm-\xp).
\end{equation}
Solving for the post-epidemic susceptible fraction yields
\begin{equation}\label{eq:xp.compute}
  \xp
  \;=\;
  -\frac{1}{\Rn}
  \Wp\!\left(-\Rn\xm e^{-\Rn\xm}\right),
\end{equation}
where $\Wp$ is the principal branch of Lambert's $W$ function
(\appref{app:LambertW}).
The proportion infected during the focal epidemic, \ie the ``true''
final size, is
\begin{equation}\label{eq:zp.compute}
\zp \;=\; \xm-\xp.
\end{equation}
Of course, calculating $\xp$ and $\zp$ in this way assumes that
transmissibility remains constant through the end of the epidemic.

Operationally, we estimate $\lamm$ from the initial exponential-growth
phase and evaluate $\Lapm$.  At a chosen time $\tau$, we reconstruct
$\Reff(\tau)$ and $Y(\tau)$ from the incidence history.
\Cref{eq:R.C.state,eq:zm.compute} then determine $\Rn$, $\xm$, and
$\zm$.  (In the simulations and Philadelphia analysis below, we use the
momentum-peak special case, for which $\Reff(\taupeak)=1$ and
$Y(\taupeak)=\ypeak$, and apply \cref{eq:R.C.lampm}.)
\Cref{eq:xp.compute} then determines $\xp$, and
\cref{eq:zp.compute} gives $\zp$.  The workflow is summarized in
\cref{fig:inference.workflow}.

\input{epimom_inference/epimom_fig_inference_workflow}

The absolute scale of the incidence curve is required because
$Y(\tau)$, unlike $\lamm$, $\Lapm$, and $\Reff(\tau)$, depends on that
scale.  Multiplying incidence by a constant multiplies $Y(\tau)$ at
every time, including its maximum $\ypeak$, by the same constant, while
leaving $\lamm$, $\Lapm$, and $\Reff(\tau)$ unchanged.  Such a
rescaling therefore changes the value of $\Rn$ inferred from
\cref{eq:R.C.state;Rn,eq:R.C.lampm}.  Incidence must therefore be
expressed as a fraction of the population, or the factor connecting the
observed time series to incidence must be known or examined explicitly,
as in the influenza mortality application below.

\beginsection{Results}{sec:results}

\beginsubsection
  {\texorpdfstring{Estimates of prior immunity and $\Rn$ from stochastic simulations}{Estimates of prior immunity and R0 from stochastic simulations}}
  {sec:est.sm.Rn.stoch}

We assessed the peak-momentum implementation of the method using
stochastic SEIR simulations for which $\Rn$ and the pre-epidemic
susceptible fraction $\xm$ were known.  The simulations spanned a wide
range of $\Rn$ for several combinations of $\xm$ and population size
$N$, allowing us to examine the effects of demographic stochasticity
on each stage of the inference.

\input{epimom_inference/epimom_fig_lamm_ypeak}
\input{epimom_inference/epimom_fig_xm_rn_estimates}

The peak-momentum implementation first estimates the initial growth
rate $\lamm$ from the rising tail and constructs the epidemic momentum
from incidence to estimate its maximum $\ypeak$.
(Details of the computation, including estimating initial conditions,
are given in \cref{app:yinit.from.incinit.finite.int.rep}.)
\Cref{fig:lamm.ypeak} shows the
relative errors in these two quantities.  Inserting the estimated
values of $\lamm$ and $\ypeak$ into \cref{eq:R.C.lampm,eq:zm.compute}
gives the corresponding estimates of $\Rn$ and $\xm$, shown
in \cref{fig:xm.R0.pred}.

In the upper panels of \cref{fig:xm.R0.pred}, grey squares show the
true values of $\xm$.  The estimated values are distributed around the
corresponding true values across the simulated parameter combinations,
with greater stochastic variation in smaller populations.  The lower
panels show the estimates of $\Rn$, which similarly lie close to the
line of perfect agreement, with accuracy improving as population size
increases.

For comparison, the smaller symbols in the lower panels show the
reproduction numbers obtained from the initial growth rate under the
assumption that the population was initially fully susceptible
($\xm=1$).  When pre-existing immunity is present, these values are
systematically smaller than the true $\Rn$, because the initial growth
rate identifies $\Reff=\Rn\xm$ rather than $\Rn$.

All simulations shown in
\cref{fig:lamm.ypeak,fig:xm.R0.pred} had equal mean latent and
infectious periods ($\ell=1$ in Eq.\,(\labelcref{eq:SEIR})); the performance of
the method was similar for other values of $\ell$.

\beginsubsection
    {\texorpdfstring{Reappraisal of the 1918 influenza pandemic in Philadelphia}{Reappraisal of the 1918 influenza pandemic in Philadelphia}}
    {sec:1918flu}

\input{epimom_inference/phila_calculations_results.tex}

Having shown that the momentum-based method recovers known values of
$\Rn$ and $\xm$ from stochastic simulations, we applied it to daily
pneumonia and influenza (P\&I) mortality recorded during the autumn
1918 main wave in Philadelphia (\cref{fig:philaflu}). Because the
mortality record spans the complete main wave, the reconstructed
momentum curve includes its peak, permitting use of the peak formula
\labelcref{eq:R.C.lampm}.

\input{epimom_inference/epimom_fig_philaflu}

Because mortality rather than incidence was reported, we used
Richardson--Lucy deconvolution to reconstruct a curve proportional to
incidence \citeref{Gold+09}.  (Details of the mortality data, the
infection-to-death distribution used for deconvolution, the
generation-interval assumptions, and the calculation of epidemic
momentum are given in the caption to \cref{fig:philaflu}.)  The
proportionality factor includes the case-fatality proportion,
$\CFPvar$, because only that fraction of infections resulted in
recorded deaths.  This unknown scale does not affect the estimated
initial growth rate \citeref{Ma+14,Earn+20,epigrowthfit}, which was
$\lamm\simeq\philalamm/\text{day}$.

Following Mills \emph{et al.}\ \citeref{Mill+04}, we assumed mean
latent and infectious periods of $\Tlat=\philaTlatdays\,$days and
$\Tinf=\philaTinfdays\,$days, respectively, hence
$\ell\equiv\Tlat/\Tinf=\philaell$.
Measuring time in units of $\Tinf$ gives $\lamm\approx\philalammdim$.
The corresponding SEIR generation-interval distribution
[\mainref{tab:models}] then yields
\begin{equation}\label{eq:phila.Lapm}
  \Lapm
  \;\approx\;
  \philaLapm.
\end{equation}
This estimate gives a lower bound on $\Rn$ that is independent of the
incidence scale: since $\xm\le1$, we have $\Rn\ge\Reff$, so
\cref{eq:lamLaplace} implies
\begin{equation}\label{eq:phila.Rn.lower}
  \Rn
  \;\geq\;
  \frac{1}{\Lapm}
  \;\approx\;
  \philaRnlowerbound.
\end{equation}

Convolving the reconstructed curve proportional to incidence with
$\Ra/\Rn$, the remaining-reproductive-potential kernel
[\cref{eq:Ra,eq:re.Y}], gives a curve proportional to epidemic
momentum.  After division by the Philadelphia population, this curve
is $\CFPvar\times Y(t)$, so its maximum is $\CFPvar \times \ypeak$.
Dividing by the representative value $\CFPvar=\philaCFPpct\%$ gives
the momentum curve $Y(t)$ shown in \cref{fig:philaflu}.
For mortality data, \cref{eq:R.C.lampm} is therefore more usefully
written
\begin{equation}\label{eq:R.C.lampm.mort}
  \Rn
  \;=\;
  \frac{\CFPvar}{\CFPvar \times \ypeak}
  \Volterra\!\Big(\frac{1}{\Lapm}\Big).
\end{equation}
The maximum of the corresponding per-capita curve before division by
$\CFPvar$ is
\begin{equation}\label{eq:phila.ymax.CFP}
  \CFPvar \times \ypeak
  \;\approx\;
  \philaypeakCFP,
\end{equation}
so that
\begin{equation}\label{eq:phila.Rn.CFP}
  \Rn
  \;\approx\;
  \philaRnoverCFP \times \CFPvar.
\end{equation}
The lower bound on $\Rn$ given in \cref{eq:phila.Rn.lower} therefore
implies a $\CFPvar$ lower bound,
\begin{equation}\label{eq:phila.CFP.lower}
\CFPvar
\;\approx\;
\frac{\Rn}{\philaRnoverCFP}
\;\gtrsim\;
\frac{\philaRnlowerbound}{\philaRnoverCFP}
\;=\;
1.6\%.
\end{equation}

Frost \citeref[p.\,593]{Frost1920} estimated that the case-fatality
proportion during the main wave of the 1918 pandemic ranged from
$0.8\%$ to $3.1\%$ across US cities.  For a group of northeastern
communities near Philadelphia, he estimated
$\CFPvar=2.05\%$ \citeref[p.\,593]{Frost1920}, consistent
with the lower bound in \cref{eq:phila.CFP.lower}.  We therefore use
$\CFPvar\simeq\philaCFPpct\%$ as a representative value for
Philadelphia.  Substitution into \cref{eq:phila.Rn.CFP} implies
\begin{equation}\label{eq:phila.Rn}
  \Rn
  \;\approx\;
  \philaRn.
\end{equation}
Equation \labelcref{eq:zm.compute} then gives
\begin{equation}\label{eq:phila.zm}
  \zm
  \;\approx\;
  \philazm,
\end{equation}
indicating that approximately $\philazmpct\%$ of the population was
immune before the main wave.

Mills \emph{et al.}\ \citeref{Mill+04} also assumed
$\CFPvar=2\%$, based on the midpoint of Frost's range
\citeref{Frost1920}, but assumed prior population immunity
$\zm=\philaMillszmpct\%$ from seasonal-influenza evidence
\citeref[pp.\,905--906]{Mill+04} (the relevant passages are reproduced
in \appref{app:quotes}).  They reported
\begin{equation}\label{eq:phila.Mills.Rn.range}
  \philaMillsRnmin
  \;\leq\;
  \Rn
  \;\leq\;
  \philaMillsRnmax
\end{equation}
for Philadelphia \citeref[SI p.\,5]{Mill+04}.  However, using the
observed initial growth rate and the SEIR generation-interval
distribution assumed both here and by Mills \emph{et al.},
$\zm=\philaMillszm$ would imply
\begin{equation}\label{eq:phila.Mills.Rn.check}
  \Rn \;=\; \frac{1}{(1-\zm)\Lapm} \;\approx\; \philaMillsRncheck,
\end{equation}
which lies outside their reported range.  Thus their stated assumption
about prior immunity and their reported estimate of $\Rn$ are not
mutually consistent under \cref{eq:zm.compute}.

Our momentum-based analysis instead estimates $\Rn$ and $\zm$
consistently from the observed initial growth and reconstructed
momentum curve.  The estimate in \cref{eq:phila.zm} suggests that a
substantial fraction (20\%) of the Philadelphia population was already
immune before the main wave, plausibly because of infection during the
spring 1918 ``herald wave''.

\beginsection{Discussion}{sec:discuss}

Initial epidemic growth does not, by itself, distinguish pathogen
transmissibility from pre-existing population immunity.  Rather, it
identifies the initial effective reproduction number $\Reff=\Rn\xm$,
the product of the basic reproduction number and the fraction of the
population susceptible before the epidemic began \citeref{McCaw+2009}.
Consequently, different combinations of $\Rn$ and $\xm$ can generate
the same initial growth.  The subsequent incidence history determines
epidemic momentum and the effective reproduction number over time;
combined with the initial-growth estimate of $\Rn\xm$, these
quantities determine $\Rn$ and $\xm$ separately, from which prior
immunity ($\zm$) and the proportion infected during the focal epidemic
($\zp$) follow.  The momentum peak need not be observed to make this
inference \labelcref{eq:R.C.state}; if it has been observed, the
simpler formula \labelcref{eq:R.C.lampm} can be used, as in our
analyses of simulations and 1918 influenza in Philadelphia
(\cref{sec:est.sm.Rn.stoch,sec:1918flu}).

The momentum-based method successfully recovered both $\Rn$ and $\xm$ from
stochastic SEIR simulations across the parameter combinations examined
(\cref{fig:lamm.ypeak,fig:xm.R0.pred}).  As expected, stochastic variation was
greater in smaller populations, but the estimates were distributed
around the corresponding true values.  By contrast, estimates obtained
from the initial growth rate while assuming $\xm=1$ were systematically
smaller than the true $\Rn$ whenever pre-existing immunity was present.
These estimates are not intrinsically inaccurate: they estimate
$\Reff=\Rn\xm$, rather than $\Rn$.  The error arises when $\Reff$ is
interpreted as the basic reproduction number despite uncertainty about
the initial susceptible fraction.

Our reappraisal of the 1918 influenza pandemic illustrates the
biological importance of this distinction.  For the main wave in
Philadelphia, which occurred in the autumn of 1918, the observed
initial growth and reconstructed epidemic momentum imply
$\Rn\approx\philaRn$ and $\zm\approx\philazm$.  Thus, approximately
$\philazmpct\%$ of the population appears to have been immune before
the main wave began.  Infection during the spring 1918 ``herald wave''
provides a plausible explanation for at least some of this immunity,
but other sources of pre-existing immunity are also possible.

The Philadelphia analysis also illustrates the consequences of
assuming prior immunity rather than estimating it.  Mills \emph{et
al.}\ \citeref{Mill+04} assumed $\zm=\philaMillszmpct\%$, but the
observed growth rate and their assumed generation-interval
distribution imply $\Rn\approx\philaMillsRncheck$, outside the range
they reported.  Thus, their assumed prior immunity and reported
estimate of $\Rn$ are not mutually consistent under the standard
relationship between epidemic growth, prior immunity, and reproduction
number given in \cref{eq:zm.compute}.  The broader point is not
specific to this earlier analysis: estimates of $\Rn$ obtained by
fixing prior immunity necessarily inherit any error in that
assumption.

\beginsubsection{Relation to previous approaches}{sec:discuss.prior}

Previous attempts to infer transmissibility and prior immunity from
epidemic time series have generally required additional approximations,
strong model assumptions, or external information.  Caley \emph{et
al.}\ \citeref{Cale+2008} estimated the time-dependent effective
reproduction number during the multi-wave 1919 influenza epidemic in
Sydney from hospitalization and mortality data.  They combined these
estimates with cumulative incidence and \emph{a priori} assumptions about periods in
which contact behaviour had returned to normal to separate susceptible
depletion from changes in contact rates.  Their analysis suggested
that more than $90\%$ of the population was initially susceptible and
gave a preferred estimate $\Rn\approx1.8$.  The inference of prior
immunity depended on the assumed clinical attack rate, the
reconstruction of the time-dependent reproduction number, and the
correct identification of periods without behavioural change.

Cope \emph{et al.}\ \citeref{Cope+2018} used approximate Bayesian
computation to fit a climate-driven stochastic epidemic model to
seasonal influenza surveillance data from New South Wales.  Their
posterior distribution showed a strong relationship between $\Rn$ and
the initial susceptible fraction: relatively high values of $\Rn$ were
associated with low susceptibility, while their product was much more
tightly constrained.  Thus, even with a substantially richer model and
observation process, the data did not sharply identify the two
quantities separately.

Bergstr{\"o}m \emph{et al.}\ \citeref{Berg+2026} proved that, in an
SIR model with under-reporting and prior immunity, the transmission
rate, reporting fraction, and initial immune fraction are not jointly
identifiable from reported incidence alone.  They showed that
identifiability can be restored by supplementing reported incidence
with an estimate of prior immunity or infection prevalence.  Their
analysis is restricted to the SIR model and requires observations
through the end of the epidemic.  By contrast, our approach, which
applies to all models represented by the \RE, uses epidemic momentum
reconstructed from the incidence history to separate $\Rn$ and $\xm$
at any time while the epidemic is in
progress \labelcref{eq:R.C.state}, even before the momentum peak has
been observed.  An unknown reporting fraction remains a limitation,
because it determines the scale of epidemic momentum and must be
estimated or constrained; in our Philadelphia analysis, the
corresponding unknown scale is the case-fatality proportion.

\beginsubsection{Data requirements and statistical challenges}{sec:discuss.limitations}

The momentum-based method requires an estimate of the intrinsic
generation-interval distribution.  This requirement is not unique to
our method: the same distribution is already needed to convert an
epidemic growth rate into an effective reproduction number
\citeref{WallLips07}.  Nevertheless, uncertainty or misspecification
in $g(\aoi)$ affects both $\Reff$ and the reconstruction of epidemic
momentum, and therefore propagates into estimates of $\Rn$ and $\xm$.
Applications should therefore be mindful of sensitivity to the assumed
form and parameters of the intrinsic generation-interval distribution.

The absolute scale of incidence is also required for point estimates of
$\Rn$ and $\xm$.  Multiplying the incidence curve by a constant leaves
its exponential growth rate unchanged but multiplies epidemic momentum
and its maximum by the same constant, thereby changing the inferred
$\Rn$.  Incidence must therefore be expressed as a fraction of the
population, or the factor connecting the observed time series to
incidence must be known or estimated.  Even when this scale factor
cannot be estimated reliably, the initial growth rate still provides a
useful lower bound on $\Rn$.  For Philadelphia,
\cref{eq:phila.Rn.lower} gives
$\Rn\gtrsim\philaRnlowerbound$, independently of the assumed
case-fatality proportion.

Mortality, hospital admissions, and other delayed observations require
reconstruction of the underlying incidence curve.  This reconstruction
depends on the relevant delay distribution and may introduce additional
uncertainty.  The unobserved incidence history before observations
begin must also be approximated, but the initial exponential phase
provides a natural extrapolation
[\appref{app:yinit.from.incinit.finite.int.rep}].

Although the general formula \labelcref{eq:R.C.state} can be evaluated
before the momentum peak, estimates based on states far from the peak
may be more sensitive to reconstruction error because both the
numerator and denominator of \cref{eq:R.C.state;Rn} decrease away from
the peak [\appref{app:peak.properties}].  Developing confidence
intervals that account for all known sources of uncertainty remains an
important statistical challenge.

\beginsubsection{Model assumptions and extensions}{sec:discuss.extensions}

The derivations in this paper assume mass-action incidence, so that
susceptibility enters incidence linearly through the susceptible
fraction of the population.  Individual differences in intrinsic
susceptibility, contact rate, and other mechanisms can instead produce
nonlinear dependence of incidence on the susceptible
fraction \citeref{WilsWorc45,Liu+86,FinkGren00,Novo08}.  In that case,
the conserved quantity and inference formulae derived here must be
modified.

\ifanonymous
The epidemic-momentum framework can be extended to incidence
that depends nonlinearly on susceptible fraction, and the
corresponding conserved quantity in this more general setting
can be derived
\citeref{epimom-nli}.
\else
We have extended the epidemic-momentum framework to incidence that
depends nonlinearly on susceptible fraction, and derived the
corresponding conserved quantity in this more general setting
\citeref{epimom-nli}.
\fi
This class provides one way to
incorporate nonlinear effects of susceptibility without explicitly
tracking population structure \cite{Novo08}.  More detailed forms of
host heterogeneity may instead require structured models that
distinguish among host types.

The present analysis assumes that transmissibility and the intrinsic
generation-interval distribution remain constant over the period
analysed.  Applications should therefore be restricted to periods over
which these assumptions are defensible.  Extending the inference method
to time-varying transmission remains a subject for future work.

Finally, the method estimates the susceptible fraction relevant to the
focal epidemic but does not identify the biological source of prior
immunity.  Such immunity may arise from previous infection with the
same pathogen, cross-reactive immunity to related strains,
vaccination, or other differences in susceptibility.  In the
Philadelphia application, infection during the spring herald wave is
one plausible source, but other sources of pre-existing immunity are
possible.

\beginsubsection{Conclusions}{sec:discuss.conclusion}

Epidemic growth reflects both pathogen transmissibility and the
immunological state of the population in which transmission occurs.
Assuming that the population was initially fully susceptible can
therefore lead to substantial underestimation of $\Rn$ and
misinterpretation of epidemic final size.  In the general \RE
framework, epidemic momentum provides a way to separate these effects
using information contained in the same epidemic time series.  Initial
growth and the subsequent incidence history yield separate estimates
of $\Rn$ and prior population immunity at any time while the epidemic
is in progress.

%% file: epimom_inference/epimom_fig_inference_workflow.tex
\begingroup
\newcommand{\workflowarrowlength}{1.5em}          
\newcommand{\workflowdecisionarrowlength}{3.0em}  
\newcommand{\workflowbranchexitlength}{1.5em}     
\newcommand{\workflowbranchoffset}{0.205\textwidth} 
\newcommand{\workflowboxwidth}{0.30\textwidth}
\newcommand{\workflowinputwidth}{0.34\textwidth}
\newcommand{\workflowdecisionwidth}{0.22\textwidth}
\newcommand{\workflowarrowheadlength}{2.2mm}
\newcommand{\workflowarrowheadwidth}{1.5mm}
\newcommand{\workflowlinewidth}{0.55pt}
\newcommand{\workflowarxivscale}{0.90}

\ifepimomarxiv
  \newcommand{\workflowfigurescale}{\workflowarxivscale}
\else
  \newcommand{\workflowfigurescale}{1}
\fi

\ifepimomarxiv
\begin{figure}[!htbp]
\else
\begin{figure*}[!t]
\fi
  \centering
  \begin{tikzpicture}[
    scale=\workflowfigurescale,
    transform shape,
    workflow process/.style={
      draw=black!65,
      rounded corners=2pt,
      fill=black!3,
      align=center,
      text width=\workflowboxwidth,
      minimum height=4.8em,
      inner sep=4pt,
      font=\fontsize{8.5}{10}\selectfont\sffamily
    },
    workflow data/.style={
      workflow process,
      trapezium,
      trapezium left angle=70,
      trapezium right angle=110,
      fill=black!1,
      text width=\workflowinputwidth,
      minimum height=5.2em
    },
    workflow decision/.style={
      diamond,
      aspect=2.15,
      draw=black!65,
      fill=black!5,
      align=center,
      text width=\workflowdecisionwidth,
      inner sep=1.5pt,
      font=\fontsize{8.5}{10}\selectfont\sffamily\bfseries
    },
    workflow branch/.style={
      workflow process,
      text width=\workflowboxwidth,
      minimum height=6.2em
    },
    workflow result/.style={
      workflow process,
      fill=black!7,
      text width=\workflowboxwidth,
      minimum height=5.2em
    },
    workflow arrow/.style={
      -{Latex[length=\workflowarrowheadlength,
              width=\workflowarrowheadwidth]},
      line width=\workflowlinewidth,
      draw=black!70
    },
    workflow line/.style={
      line width=\workflowlinewidth,
      draw=black!70
    },
    workflow label/.style={
      fill=white,
      inner sep=1.5pt,
      font=\fontsize{8}{9}\selectfont\sffamily\bfseries
    }
  ]
    \node[workflow data] (inputs) {
      \textbf{Inputs}\\[2pt]
      Infection history $\inc(\tau)$\\[2pt]
      Generation-interval distribution $g(\aoi)$
    };

    \node[workflow process,below=\workflowarrowlength of inputs] (growth) {
      \textbf{Initial growth}\\[2pt]
      Estimate $\lamm$ and evaluate $\Lapm$\\[2pt]
      \Cref{eq:Reff.initial}
    };

    \node[workflow process,below=\workflowarrowlength of growth] (momentum) {
      \textbf{Epidemic momentum}\\[2pt]
      Reconstruct $Y(\tau)$ over the available interval\\[2pt]
      \Cref{eq:re.Y.finite}
    };

    \node[workflow decision,below=\workflowarrowlength of momentum] (decision) {
      Does the reconstructed\\momentum curve include\\its peak?
    };

    \node[workflow branch,below=\workflowdecisionarrowlength of decision,
          xshift=-\workflowbranchoffset] (peak) {
      \textbf{Peak special case}\\[2pt]
      At $\taupeak$, use $Y(\taupeak)=\ypeak$ and
      $\Reff(\taupeak)=1$\\[2pt]
      \Cref{eq:R.C.lampm}
    };

    \node[workflow branch,below=\workflowdecisionarrowlength of decision,
          xshift=\workflowbranchoffset] (general) {
      \textbf{General state formula}\\[2pt]
      Choose $\tau$ with $Y(\tau)>0$ and reconstruct $\Reff(\tau)$\\[2pt]
      \Cref{eq:re.Reff,eq:R.C.state}
    };

    \coordinate (branchbottom) at (decision.center |- peak.south);
    \coordinate[below=\workflowbranchexitlength of branchbottom] (merge);

    \node[workflow result,below=\workflowarrowlength of merge] (inference) {
      \textbf{Joint inference}\\[2pt]
      Determine $\Rn$, $\xm$, and $\zm=1-\xm$\\[2pt]
      \Cref{eq:R.C.state,eq:R.C.lampm,eq:zm.compute}
    };

    \node[workflow process,below=\workflowarrowlength of inference] (finalsize) {
      \textbf{Final size}\\[2pt]
      Calculate $\xp$ and $\zp=\xm-\xp$\\[2pt]
      \Cref{eq:xp.compute,eq:zp.compute}
    };

    \draw[workflow arrow] (inputs) -- (growth);
    \draw[workflow arrow] (growth) -- (momentum);
    \draw[workflow arrow] (momentum) -- (decision);

    \draw[workflow arrow] (decision.south west) --
      node[workflow label,pos=0.25,below left] {Yes} (peak.north);
    \draw[workflow arrow] (decision.south east) --
      node[workflow label,pos=0.25,below right] {No} (general.north);

    \draw[workflow line] (peak.south) |- (merge);
    \draw[workflow line] (general.south) |- (merge);
    \draw[workflow arrow] (merge) -- (inference.north);
    \draw[workflow arrow] (inference) -- (finalsize);
  \end{tikzpicture}
  \caption{\textbf{Workflow for jointly estimating transmissibility
  and prior immunity.}  Infection history $\inc(\tau)$ and the
  intrinsic generation-interval distribution $g(\aoi)$ determine the
  initial effective reproduction number $\Reff(-\infty)$ (which is
  equal to the product $\Rn\xm$) and the time course of the epidemic
  momentum $Y(\tau)$.  If the reconstructed momentum curve includes
  its peak value $\ypeak$, then the peak special case is used;
  otherwise, the general state formula can be evaluated using
  reconstructed $\Reff(\tau)$ at any time for which $Y(\tau)>0$.
  Applying the workflow assumes that $g(\aoi)$ is known and that
  transmissibility is constant from initial growth through the chosen
  state.  Calculation of final-size quantities additionally assumes
  that transmissibility remains constant through the epidemic
  endpoint.}  \label{fig:inference.workflow}
\ifepimomarxiv
\end{figure}
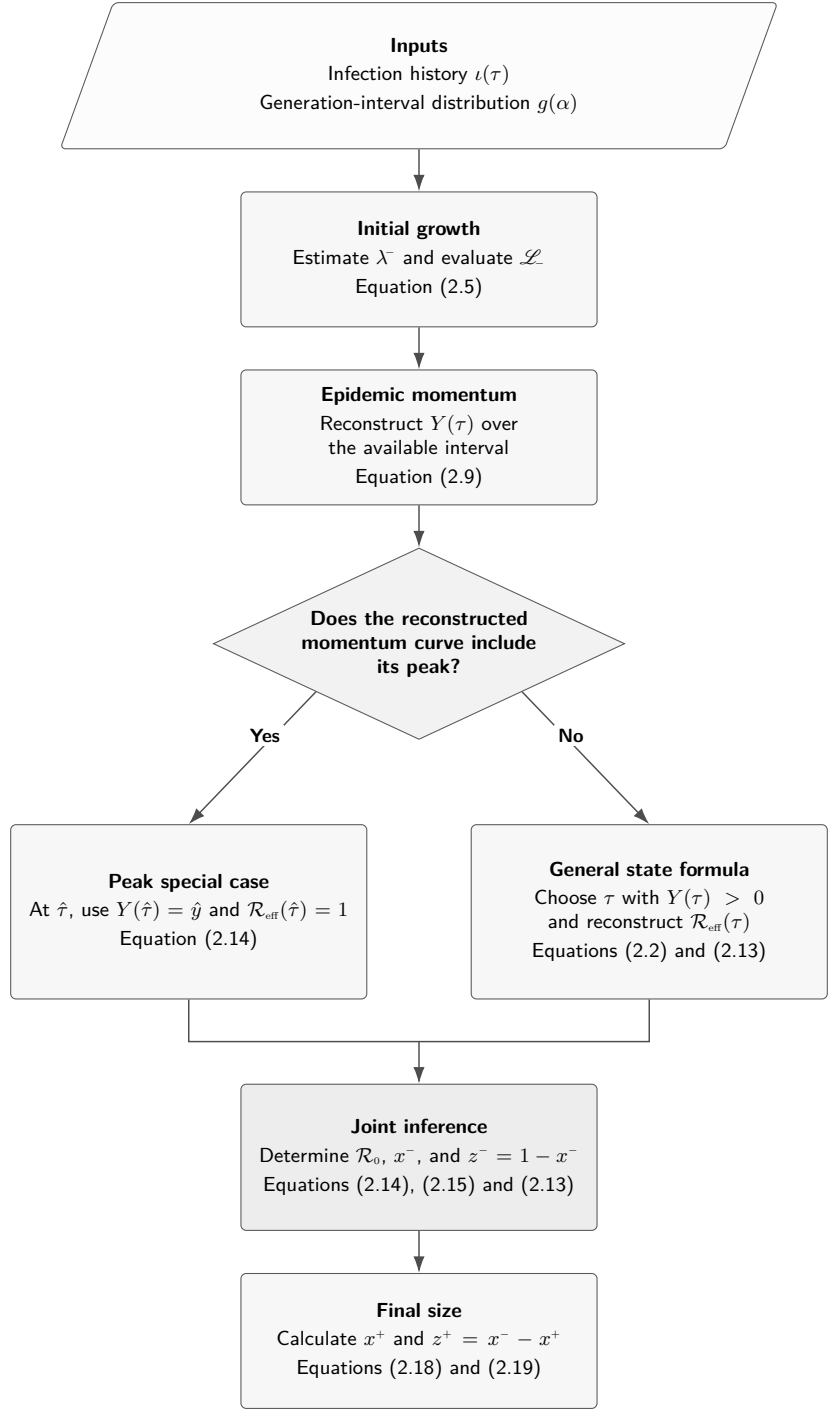
\else
\end{figure*}
\fi
\endgroup

%% file: epimom_inference/epimom_fig_lamm_ypeak.tex
\begin{figure*}
\begin{center}
\includegraphics[width=0.95\textwidth]{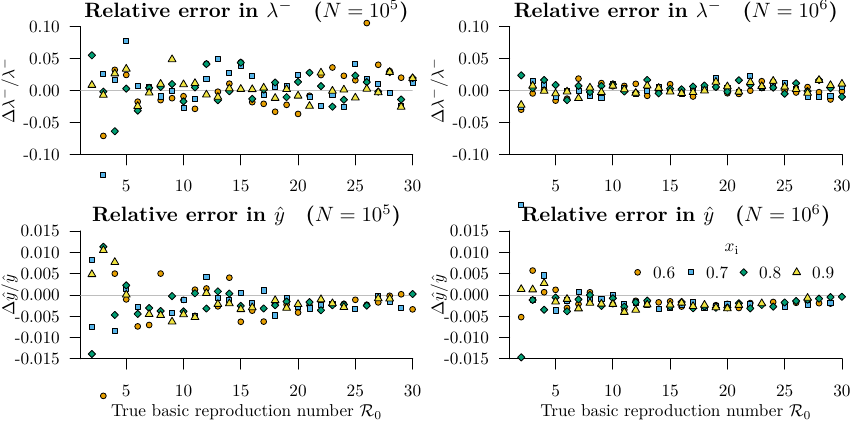}
\end{center}
\caption{\textbf{Estimates of initial growth rate $\lamm$ and peak
    epidemic momentum $\ypeak$ from stochastic SEIR simulations.}  The
  top panels show the relative error in the initial growth rate
  $\lamm$ as a function of the true $\Rn$ specified in the simulations
  (with population size $N=10^5$ on the left and $N=10^6$ on the
  right).  Symbols and colours indicate the initial susceptible
  proportion $\xinit$.  The exact value of $\lamm$ is given by the
  SEIR row in \protect\mainref{tab:models}.
Simulations were carried out with equal mean latent and infectious
periods ($\ell=1$) and incidence time series were obtained by
``observing'' five times per infectious period, corresponding to daily
data for a disease with a five-day infectious period.
Realistic initial conditions were chosen by ensuring
approximate agreement with the initial exponential growth phase of the
SEIR ODEs.
Estimates of $\lamm$ were obtained by applying the \texttt{R} package
\textsf{epigrowthfit} \citeref{Ma+14,Earn+20,epigrowthfit} to the simulated
incidence time series.
The second row of panels shows the relative error in the peak epidemic
momentum ($\ypeak$, the exact value of which is given by $Y(\xpeak)$
in \protect\theoryref{eq:Yofx}).  Epidemic momentum $Y(\tau)$ was
estimated from the simulated cumulative incidence $\cuminc(\tau)$
and the generation-interval distribution $g(\aoi)$ using
\protect\cref{eq:re.Y.g,tab:models}.
A small systematic underestimate in $\ypeak$ is evident, but the
magnitude of the relative error in $\ypeak$ is an order of magnitude
smaller than the magnitude of the relative error in $\lamm$, so the
systematic error has a negligible effect on the estimate of $\Rn$.
}
\label{fig:lamm.ypeak}
\end{figure*}

%% file: epimom_inference/epimom_fig_xm_rn_estimates.tex
\begin{figure*}
\newcommand{\bigfigwidth}{\textwidth}
\begin{center}
  \null\hspace{0cm}\includegraphics[width=\bigfigwidth]{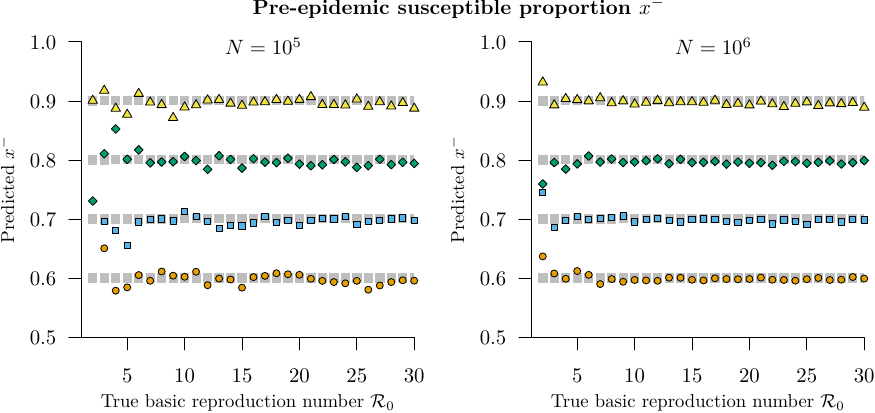}
  \vspace{20pt}
  \includegraphics[width=\bigfigwidth]{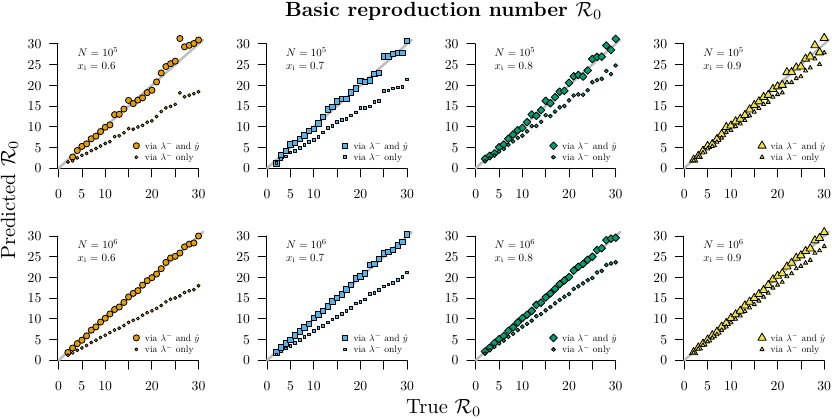}
\end{center}
\caption{\textbf{Prior population immunity ($\zm=1-\xm$)
and basic reproduction number ($\Rn$) estimated from stochastic SEIR simulations.}
Exploiting the epidemic momentum, we successfully disentangle and
accurately estimate both $\zm$ and $\Rn$.
The top panels show the predicted pre-epidemic susceptible proportion
$\xm$, so the pre-existing level of population immunity is $\zm=1-\xm$
[estimated via \protect\cref{eq:zm.compute}].  The true $\xm$ associated with
the deterministic skeleton of the model [computed via \cref{eq:zm.compute} using 
the exact value of $\lamm$ from \cref{tab:models}]
is indicated with grey squares.
Symbols and colours are associated with the initial susceptible
proportion $\xinit$ as in \protect\cref{fig:lamm.ypeak}.
The bottom panels show the predicted $\Rn$ from the same simulations.
The smaller symbols show the Wallinga--Lipsitch \citeref{WallLips07}
estimate obtained from the estimated growth rate $\lamm$ and the
assumed generation-interval distribution $g(\aoi)$; interpreting this
quantity as $\Rn$ assumes that $\xm=1$.  When $\xm<1$, the quantity
estimated is instead the initial effective reproduction number
$\Reff(-\infty)=\Rn\xm$.  The larger symbols show $\Rn$ estimated from
$\lamm$, $g(\aoi)$, and the estimated peak epidemic momentum $\ypeak$
using \protect\cref{eq:R.C.lampm}.  The grey line corresponds to
``Predicted $\Rn=$ True $\Rn$''.}
\label{fig:xm.R0.pred}
\end{figure*}

%% file: epimom_inference/phila_calculations_results.tex
\newcommand{\philaTlatdays}{1.9}
\newcommand{\philaTinfdays}{4.1}
\newcommand{\philaell}{0.463}
\newcommand{\philalamm}{0.16}
\newcommand{\philalammdim}{0.655}
\newcommand{\philaLapm}{0.464}
\newcommand{\philaRnlowerbound}{2.16}
\newcommand{\philaypeakCFP}{0.00287}
\newcommand{\philaCFP}{0.02}
\newcommand{\philaCFPpct}{2}
\newcommand{\philaCFPminpct}{0.739}
\newcommand{\philaCFPlowerbound}{0.0159}
\newcommand{\philaCFPlowerboundpct}{1.59}
\newcommand{\philaypeak}{0.143}
\newcommand{\philaxm}{0.797}
\newcommand{\philazm}{0.203}
\newcommand{\philazmpct}{20}
\newcommand{\philaRnoverCFP}{135}
\newcommand{\philaRn}{2.71}
\newcommand{\philaincpeakdate}{29 September 1918}
\newcommand{\philaypeakdate}{1 October 1918}
\newcommand{\philapimpeakdate}{11 October 1918}
\newcommand{\philapimpeakvalue}{803}
\newcommand{\philalastdate}{17 December 1918}
\newcommand{\philaMillszm}{0.3}
\newcommand{\philaMillszmpct}{30}
\newcommand{\philaMillsxm}{0.7}
\newcommand{\philaMillsRnmin}{1.7}
\newcommand{\philaMillsRnmax}{2.4}
\newcommand{\philaMillsRncheck}{3.08}

%% file: epimom_inference/epimom_fig_philaflu.tex
\begin{figure*}[!t]
  \centering
  \includegraphics[width=0.75\textwidth]{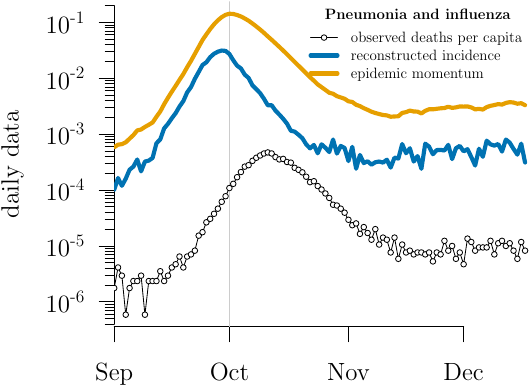}
\caption{\textbf{1918 influenza pandemic in Philadelphia, USA.}  Daily
  deaths from pneumonia and influenza (P\&I) were recorded from 1
  September to 31 December 1918 \citeref{Roge20}.  We deconvolved the
  observed mortality time series to reconstruct a curve proportional
  to daily incidence, using an empirically estimated infection-to-death
  distribution: as detailed in previous work \citeref{Gold+09} and
  implemented in the \textsf{fastbeta} R package
  \citeref{Jaga+20,fastbeta}, gamma distributions were fitted to an
  empirical incubation-period distribution \citeref[Figure~1]{Mose+79}
  and an empirical symptom-onset-to-death distribution
  \citeref[Chart~2]{KeetCush18}, which were then convolved to obtain
  the infection-to-death distribution.  We then used $\Ra/\Rn$, the
  fraction of reproductive potential remaining at infection age
  $\aoi$ [\protect\cref{eq:Ra,eq:re.Y}], to calculate a curve
  proportional to epidemic momentum.  The ratio $\Ra/\Rn$ was obtained
  from the SEIR generation-interval distribution $g(\aoi)$ in
  \protect\cref{tab:models}, with $\ell=\philaell$.  
  The estimated incidence $\inc(t)$ and epidemic momentum $Y(t)$ shown
  here were obtained by dividing the reconstructed curves by the
  Philadelphia population and by the representative case-fatality
  proportion $\CFPvar=\philaCFPpct\%$ discussed in
  \protect\cref{sec:1918flu}.
  The peak of the observed daily P\&I mortality occurred on
  \philapimpeakdate\ (with \philapimpeakvalue\ P\&I deaths), whereas
  estimated incidence peaked on \philaincpeakdate\ and estimated
  epidemic momentum peaked on \philaypeakdate\ (vertical grey line).
  Note that peak momentum always occurs after peak incidence (see
  \protect\appref{app:peak.properties}).  Associated estimates of
  $\Rn$ and population immunity are discussed in the main text in
  \protect\cref{sec:1918flu}.}
  \label{fig:philaflu}
\end{figure*}

%% file: epimom_inference/epimom_inference_appendices.tex
\input{epimom_inference/epimom_app_finite_inc_history.tex} 
\input{epimom_inference/epimom_app_peak_properties} 

\begingroup
\newcommand{\inferenceLambertWNote}{%
  This appendix is identical to the corresponding appendix in
  \theoryref{app:LambertW} and is included here for convenience.}
\input{shared/epimom_app_lambertw}
\endgroup

\input{epimom_inference/epimom_app_quotes} 

\input{epimom_inference/epimom_table_models} 

%% file: epimom_inference/epimom_app_finite_inc_history.tex
\beginappendix{Epidemic momentum from finite incidence history}{app:yinit.from.incinit.finite.int.rep}
\appfilename{epimom_inference/epimom_app_finite_inc_history.tex}
\appbackrefs{%
  \appbackref{finite-history-momentum}{main text: finite-history momentum calculation}%
}

For $\tau\geq\tauinit$, splitting \cref{eq:re.Y} over infections that
occurred before and after observations began gives
\begin{equation}
  Y(\tau)
  \;=\;
  \int_{\tau-\tauinit}^{\infty}
    \inc(\tau-\aoi)
    \frac{\Ra}{\Rn}\,\dee\aoi
  +
  \int_0^{\tau-\tauinit}
    \inc(\tau-\aoi)
    \frac{\Ra}{\Rn}\,\dee\aoi.
  \label{eq:re.Y.finite.split}
\end{equation}
The second integral is computed directly from the observed incidence
history.  During the initial exponential-growth phase,
\cref{eq:inc.exp}, with $\inc(\tauinit)=\incinit$, gives
\begin{equation}
  \inc(\tau-\aoi)
  \;\approx\;
  \incinit
  e^{\lamm(\tau-\tauinit)}
  e^{-\lamm\aoi},
  \qquad
  \aoi>\tau-\tauinit,
  \label{eq:re.inc.prehistory.approx}
\end{equation}
where the initial growth rate $\lamm$ is also estimated from
the observed incidence time series \citeref{Ma+14,Earn+20,epigrowthfit}.
The first integral in \cref{eq:re.Y.finite.split} is therefore
approximated by
\begin{equation}
  \incinit e^{\lamm(\tau-\tauinit)}
  \int_{\tau-\tauinit}^{\infty}
    e^{-\lamm\aoi}
    \frac{\Ra}{\Rn}\,\dee\aoi.
  \label{eq:re.Y.prehistory.approx}
\end{equation}

Using the definition of $\Ra$ in \cref{eq:Ra}, we obtain
\ifdefined\epimomnarrowcolumns
\begin{align}
  &\int_{\tau-\tauinit}^{\infty}
    e^{-\lamm\aoi}
    \frac{\Ra}{\Rn}\,\dee\aoi
  \notag\\
  &\;=\;
  \int_0^\infty
    e^{-\lamm(\aoi+\tau-\tauinit)}
    \int_{\aoi+\tau-\tauinit}^{\infty}
      g(\aoidum)\,\dee\aoidum\,\dee\aoi
  \notag\\
  &\;=\;
  e^{-\lamm(\tau-\tauinit)}
  \int_0^\infty
    e^{-\lamm\aoi}
    \int_\aoi^\infty
      g(\aoidum+\tau-\tauinit)\,\dee\aoidum\,\dee\aoi
  \notag\\
  &\;=\;
  e^{-\lamm(\tau-\tauinit)}
  \int_0^\infty
    \int_0^{\aoidum}
      e^{-\lamm\aoi}\,\dee\aoi\,
    g(\aoidum+\tau-\tauinit)\,\dee\aoidum
  \notag\\
  &\;=\;
  \frac{e^{-\lamm(\tau-\tauinit)}}{\lamm}
  \int_0^\infty
    \big(1-e^{-\lamm\aoidum}\big)
    g(\aoidum+\tau-\tauinit)\,\dee\aoidum
  \notag\\
  &\;=\;
  \frac{e^{-\lamm(\tau-\tauinit)}}{\lamm}
  \left[
    \Lap{g_{\tau-\tauinit}}(0)
    -
    \Lap{g_{\tau-\tauinit}}(\lamm)
  \right],
  \label{eq:Yinc.const}
\end{align}
\else
\begin{align}
  \int_{\tau-\tauinit}^{\infty}
    e^{-\lamm\aoi}
    \frac{\Ra}{\Rn}\,\dee\aoi
  &\;=\;
  \int_0^\infty
    e^{-\lamm(\aoi+\tau-\tauinit)}
    \int_{\aoi+\tau-\tauinit}^{\infty}
      g(\aoidum)\,\dee\aoidum\,\dee\aoi
  \notag\\
  &\;=\;
  e^{-\lamm(\tau-\tauinit)}
  \int_0^\infty
    e^{-\lamm\aoi}
    \int_\aoi^\infty
      g(\aoidum+\tau-\tauinit)\,\dee\aoidum\,\dee\aoi
  \notag\\
  &\;=\;
  e^{-\lamm(\tau-\tauinit)}
  \int_0^\infty
    \int_0^{\aoidum}
      e^{-\lamm\aoi}\,\dee\aoi\,
    g(\aoidum+\tau-\tauinit)\,\dee\aoidum
  \notag\\
  &\;=\;
  \frac{e^{-\lamm(\tau-\tauinit)}}{\lamm}
  \int_0^\infty
    \big(1-e^{-\lamm\aoidum}\big)
    g(\aoidum+\tau-\tauinit)\,\dee\aoidum
  \notag\\
  &\;=\;
  \frac{e^{-\lamm(\tau-\tauinit)}}{\lamm}
  \left[
    \Lap{g_{\tau-\tauinit}}(0)
    -
    \Lap{g_{\tau-\tauinit}}(\lamm)
  \right],
  \label{eq:Yinc.const}
\end{align}
\fi
where
\begin{equation}
  g_{\tau-\tauinit}(\aoi)
  \;=\;
  g(\aoi+\tau-\tauinit)
  \label{eq:re.g.shifted}
\end{equation}
is the shifted generation-interval kernel, with Laplace transform
\begin{equation}
  \Lap{g_{\tau-\tauinit}}(\lambda)
  \;=\;
  \int_0^\infty
    e^{-\lambda\aoi}
    g(\aoi+\tau-\tauinit)\,\dee\aoi.
  \label{eq:re.g.shifted.Laplace}
\end{equation}

Substituting \cref{eq:Yinc.const} into
\cref{eq:re.Y.prehistory.approx,eq:re.Y.finite.split} gives the
finite-history representation
\ifdefined\epimomnarrowcolumns
\begin{equation}
    \begin{aligned}
      Y(\tau)
      &\;\approx\;
      \frac{\incinit}{\lamm}
      \left[
        \Lap{g_{\tau-\tauinit}}(0)
        -
        \Lap{g_{\tau-\tauinit}}(\lamm)
      \right]
      \\
      &\qquad
      +
      \int_0^{\tau-\tauinit}
        \inc(\tau-\aoi)
        \frac{\Ra}{\Rn}\,\dee\aoi.
    \end{aligned}
  \label{eq:re.Y.finite.Laplace}
\end{equation}
\else
\begin{equation}
    Y(\tau)
    \;\approx\;
    \frac{\incinit}{\lamm}
    \left[
      \Lap{g_{\tau-\tauinit}}(0)
      -
      \Lap{g_{\tau-\tauinit}}(\lamm)
    \right]
    +
    \int_0^{\tau-\tauinit}
      \inc(\tau-\aoi)
      \frac{\Ra}{\Rn}\,\dee\aoi.
  \label{eq:re.Y.finite.Laplace}
\end{equation}
\fi
The first term reconstructs the contribution from infections that
occurred before observations began, while the second is calculated
directly from the observed incidence history.

For the SEIR generation-interval distribution given in
\cref{tab:models}, the shifted Laplace transform is
\ifdefined\epimomnarrowcolumns
\begin{equation}
  \Lap{g_{\tau-\tauinit}}(\lambda)
  \;=\;
  \begin{cases}
    \displaystyle
    \frac{1}{1-\ell}
    \left[
      \begin{aligned}
        &\frac{e^{-(\tau-\tauinit)}}{1+\lambda}
        \\
        &\quad-
        \frac{\ell e^{-(\tau-\tauinit)/\ell}}
             {1+\ell\lambda}
      \end{aligned}
    \right],
    & \ell\ne1,
    \\[7ex]
    \displaystyle
    \frac{
      1+(1+\lambda)(\tau-\tauinit)
    }{
      (1+\lambda)^2
    }
    e^{-(\tau-\tauinit)},
    & \ell=1.
  \end{cases}
  \label{eq:re.g.shifted.Laplace.SEIR}
\end{equation}
\else
\begin{equation}
  \Lap{g_{\tau-\tauinit}}(\lambda)
  \;=\;
  \begin{cases}
    \displaystyle
    \frac{1}{1-\ell}
    \left[
      \frac{e^{-(\tau-\tauinit)}}{1+\lambda}
      -
      \frac{\ell e^{-(\tau-\tauinit)/\ell}}
           {1+\ell\lambda}
    \right],
    & \ell\ne1,
    \\[3ex]
    \displaystyle
    \frac{
      1+(1+\lambda)(\tau-\tauinit)
    }{
      (1+\lambda)^2
    }
    e^{-(\tau-\tauinit)},
    & \ell=1.
  \end{cases}
  \label{eq:re.g.shifted.Laplace.SEIR}
\end{equation}
\fi
It follows that the contribution from the unobserved incidence history
is
\ifdefined\epimomnarrowcolumns
\begin{equation}
    \begin{aligned}
      &\frac{\incinit}{\lamm}
      \left[
        \Lap{g_{\tau-\tauinit}}(0)
        -
        \Lap{g_{\tau-\tauinit}}(\lamm)
      \right]
      \\
      &\qquad\;=\;
      \begin{cases}
        \displaystyle
        \frac{\incinit}{1-\ell}
        \left[
          \begin{aligned}
            &\frac{e^{-(\tau-\tauinit)}}{1+\lamm}
            \\
            &\quad-
            \frac{\ell^2e^{-(\tau-\tauinit)/\ell}}
                 {1+\ell\lamm}
          \end{aligned}
        \right],
        & \ell\ne1,
        \\[7ex]
        \displaystyle
        \incinit e^{-(\tau-\tauinit)}
        \frac{
          2+(\tau-\tauinit)
          +\lamm\big[1+(\tau-\tauinit)\big]
        }{
          (1+\lamm)^2
        },
        & \ell=1.
      \end{cases}
    \end{aligned}
  \label{eq:re.Y.prehistory.SEIR}
\end{equation}
\else
\begin{equation}
    \frac{\incinit}{\lamm}
    \left[
      \Lap{g_{\tau-\tauinit}}(0)
      -
      \Lap{g_{\tau-\tauinit}}(\lamm)
    \right]
    \;=\;
    \begin{cases}
      \displaystyle
      \frac{\incinit}{1-\ell}
      \left[
        \frac{e^{-(\tau-\tauinit)}}{1+\lamm}
        -
        \frac{\ell^2e^{-(\tau-\tauinit)/\ell}}
             {1+\ell\lamm}
      \right],
      & \ell\ne1,
      \\[3ex]
      \displaystyle
      \incinit e^{-(\tau-\tauinit)}
      \frac{
        2+(\tau-\tauinit)
        +\lamm\big[1+(\tau-\tauinit)\big]
      }{
        (1+\lamm)^2
      },
      & \ell=1.
    \end{cases}
  \label{eq:re.Y.prehistory.SEIR}
\end{equation}
\fi
These expressions allow the contribution from infections that occurred
before $\tauinit$ to be calculated directly in the SEIR examples,
without numerical integration over an unobserved incidence history.

%% file: epimom_inference/epimom_app_peak_properties.tex
\beginappendix{Properties of peak momentum}{app:peak.properties}

When the momentum peak has been observed, evaluating
\cref{eq:R.C.state} there has several advantages.  First, the peak
condition fixes $\Reff(\taupeak)=1$, so \cref{eq:R.C.lampm} does not
require a separate estimate of $\Reff(\taupeak)$.  Second, the Volterra
function is stationary at one:
\begin{equation}
  \Volterra'(1)
  \;=\;
  0,
  \qquad
  \Volterra(1+\epsilon)
  \;=\;
  \frac{\epsilon^2}{2}
  +O(\epsilon^3).
  \label{eq:Volterra.peak.expansion}
\end{equation}
Consequently, a small error that replaces $\Reff(\taupeak)=1$ by
$1+\epsilon$ changes the numerator of \cref{eq:R.C.state;Rn} only at
second order.

\Cref{eq:CQ;Reff} shows that the difference of Volterra functions in
the numerator of \cref{eq:R.C.state;Rn} equals $\Rn Y(\tau)$.
The numerator and denominator of \cref{eq:R.C.state;Rn} are
therefore both maximal at $\taupeak$ and decrease on either side,
approaching zero in the initial and final epidemic tails.  For
reconstruction errors of comparable absolute magnitude, their smaller
values away from the peak make their ratio more sensitive to errors.
Once the peak has been reliably reconstructed, evaluating the
estimator \labelcref{eq:R.C.state;Rn} at a later state again requires
$\Reff(\tau)$ while the numerator and $Y(\tau)$ both decrease, and
therefore offers no obvious advantage over \cref{eq:R.C.lampm}.

Finally, if $Y$ is smooth, then $Y'(\taupeak)=0$, and a small timing
error $\delta$ gives
\begin{equation}
  Y(\taupeak+\delta)
  \;=\;
  \ypeak
  +\frac{1}{2}Y''(\taupeak)\delta^2
  +O(\delta^3).
  \label{eq:Y.peak.expansion}
\end{equation}
Evaluating the smooth momentum curve at a nearby time therefore
changes its value only at second order.  These calculations do not
show that the peak estimator \labelcref{eq:R.C.lampm} minimizes bias
or variance under a particular observation model.  Estimating a
maximum from noisy data can itself introduce uncertainty or selection
bias \cite{DaveNich20}, and errors in the initial growth rate,
incidence calibration, and generation-interval distribution remain.
Observations after the peak may improve reconstruction or smoothing of
the peak \cite[pp.\,50--51, 354]{Bolk08}, help quantify uncertainty,
or help assess the model assumptions.

Epidemic momentum always peaks after incidence.
To see this, let $\tauincmax$ denote the time at which the incidence
curve attains its global maximum.
Using
\cref{eq:state.dynamics;Y.aoi} and the fact that $g$ integrates to one,
we have
\begin{align}
  \left.\ddtau{Y}\right|_{\tau=\tauincmax}
  &\;=\;
  \inc(\tauincmax)
  -
  \int_0^\infty
    \inc(\tauincmax-\aoi)g(\aoi)\,\dee\aoi
  \notag
  \\
  &\;=\;
  \int_0^\infty
    \bigl[\inc(\tauincmax)
    -\inc(\tauincmax-\aoi)\bigr]
    g(\aoi)\,\dee\aoi
  \;\geq\;
  0.
  \label{eq:re.Y'}
\end{align}
The inequality is strict whenever $g$ assigns positive weight to
infection ages for which incidence was below its maximum.  Momentum
therefore continues to increase at the incidence peak under this
condition.  Peak momentum always occurs at $\xpeak$
\theoryref{sec:epimom}, whereas \cref{eq:re;X} shows that the fraction
susceptible is monotone decreasing.  Hence the fraction susceptible at
peak incidence exceeds $\xpeak$.

%% file: shared/epimom_app_lambertw.tex
\beginappendix{\texorpdfstring{Lambert's $W$-function}{Lambert's W-function}}{app:LambertW}

\ifdefined\inferenceLambertWNote
  \inferenceLambertWNote
\fi

\appfilename{shared/epimom_app_lambertw.tex}
\appbackrefs{\appbackref{lambert-w}{Lambert $W$}}

If $\Winv(z) = ze^{z}$, Lambert's
$W$-function $W(z)$ 
(\citeref{Corl+96};
\citeref[\S4.13]{NIST:DLMF})
is the multivalued inverse of $\Winv$.  Its branches are denoted
$W_k(z)$, for integer $k$, and satisfy $\Winv(W_k(z))=z$ on their
respective domains.  We use the two real
branches: $W_{-1}$ maps $[-\frac{1}{e},0)$ to $(-\infty,-1]$, and
$W_{0}$ maps $[-\frac{1}{e},\infty)$ to $[-1,\infty)$.
Conversely, each branch $W_k$ is the inverse of $\Winv$ restricted to
the range of that branch.  In particular, on the real line, the
corresponding partial inverse relations are
\begin{subequations}\label{eq:Wid}
	\begin{align}
		W_{-1}(\Winv(z)) &\;=\; z \quad \text{if $z \leq -1$}\\
		W_{0}(\Winv(z)) &\;=\; z \quad \text{if $z \geq -1$}.
	\end{align}
\end{subequations}
While the standard notation $W_k$ is chosen to indicate
the winding number associated with the given branch, for our purposes
it is more convenient to write $\Wm$ for $W_{-1}$ and $\Wp$ for $W_0$,
so we can write expressions involving $\Wpm$, where the $\pm$ matches
the corresponding sign in $\xpm$ and/or $\lampm$ ($\Wp$ and $\Wm$ are 
also written ${\rm Wp}$ and ${\rm Wm}$ \citeref[\S4.13]{NIST:DLMF}).

%% file: epimom_inference/epimom_app_quotes.tex
\beginappendix{Previous estimates of prior immunity to 1918 influenza}{app:quotes}

{\bfseries Mills \emph{et al.}\ \citeref[pp.\,905--906]{Mill+04} state:} ``The proportion of the population susceptible at the start of the pandemic determines the relationship between $R$ and the basic reproductive number ($\Rn$), which is the number of secondary cases generated by a primary case in a completely susceptible population$^2$.  Frost hypothesized that a 1918 pandemic-like strain spread throughout America in the spring of 1918 (ref. 22), and recent analyses support this ‘herald wave’ hypothesis$^{23}$. Anecdotal evidence suggests that those who fell ill in the spring were protected from disease in the autumn pandemic$^{24}$. Nevertheless, a large majority of the population was probably susceptible to the A/H1N1 pandemic strain in September 1918. In a typical epidemic transmission season, 15–25\% of the population becomes infected with influenza$^4$.  The herald wave is believed to have arrived late in the 1917–18 transmission season. Using 70\% as a conservative lower bound for the fraction susceptible at the start of the autumn pandemic, the medians for our initial and extreme $\Rn$ are 2.9 and 3.9.''
 
\bigskip
\noindent
{\bfseries Frost \citeref[p.\,593]{Frost1920} states:} ``The case fatality, or ratio of deaths to total cases of influenza, varied in the localities surveyed from 3.1 per cent in New London to 0.8 per cent in San Antonio, the variations showing no consistent relation to incidence rates. There is, however, some relation to geographic location, namely, that the highest case-fatality rates occurred in New London, San Francisco, Baltimore, and minor Maryland communities, in the order named––that is, in communities representing, respectively, the northern half of the Atlantic seaboard and the Pacific coast.  In the central and southern cities the case fatality was generally notably lower. Combining the eleven localities into three groups comprising, respectively––(1) San Francisco, (2) New London, Baltimore, and minor Maryland communities, (3) central and southern cities, comprising all other localities, the case-fatality rates in these three groups are, respectively, 2.33, 2.05, and 1.08 per cent. This is of interest in connection with the observation that from the standpoint of mortality rates the epidemic was generally more severe along the northern Atlantic Seaboard and the Pacific Coast than in the Central States.''

%% file: epimom_inference/epimom_table_models.tex
\newcommand{\modeltablerule}{\rule{\linewidth}{1.0pt}}
\newcommand{\modeltablescaleby}{0.7}
\newcommand{\mysqueeze}{\vspace{-20pt}}
\newcommand{\mymedsqueeze}{\vspace{-18pt}}
\newcommand{\mysmallsqueeze}{\vspace{-10pt}}

\newcommand{\tablecaptiontitle}{{\bfseries\slshape Standard infectious disease transmission models.}}

\newcommand{\tablecaptioncontent}{\mdseries
The susceptible-infectious-removed (\textbf{SIR}) model, first
proposed by Kermack and McKendrick \citeref{KermMcKe27}, assumes that all infected
individuals are equally infectious, and immunity upon recovery is
permanent.  It is represented with two equations in standard form
[(\protect\hypercref{eq:SIRraw}), with parameters $\beta$, the transmission
  rate, $\gamma$, the removal rate, and population size $N$]
or dimensionless form
[(\protect\hypercref{eq:SIR}), with parameter $\Rn$, the basic reproduction
  number, and time measured in units of the mean infectious period,
  $\Tinf=\gamma^{-1}$].  The generation interval
distribution is identical to the infectious period distribution, so
the mean generation interval is $\gimean = \gamma^{-1}$.
In this simple model, the epidemic momentum is equal to the prevalence.

\medskip
Most infectious diseases have a non-negligible \emph{latent period},
\ie there is a delay between initial infection and becoming
infectious.  The susceptible-exposed-infectious-removed
(\textbf{SEIR}) model introduces an \emph{exposed} stage (E) of mean
duration $\Erate^{-1}$, when individuals are infected but not yet
infectious \citeref{AndeMay91}.  The mean generation interval $\gimean$
is the sum of the means of the latent and infectious periods
\citeref{Sven07,ChamDush15}.  In dimensionless units, we write the mean
latent period as $\ell$, expressed as a proportion of the mean infectious
period, so the mean generation interval is $\gimean=\ell+1$ in these
units.  The standard form is (\protect\hypercref{eq:SEIRraw}) and the dimensionless
form is (\protect\hypercref{eq:SEIR}).
We denote the proportions susceptible, exposed, and infectious by $X$,
$\Eprop$, and $\Iprop$, respectively, and---as in the SIR model---the
epidemic momentum $Y$ corresponds to the \emph{total proportion
infected}, \ie $Y=\Eprop+\Iprop$ (see \theoryref{sec:SEIR}),
consistent with our notation for the SIR model (\hypercref{eq:SIR}).
The \emph{per capita} rates at which individuals leave the exposed and
infectious compartments are $\Erate$ and $\Irate$, respectively.  
The basic reproduction
number is $\Rn=\beta/\Irate$ and the mean latent period (as a
proportion of the mean infectious period $\Irate^{-1}$) is
$\ell=\Irate/\Erate$.

\medskip
Generic epidemic models can be specified using the \textbf{\RE}, which
relates the susceptible fraction $X$ to the
force of infection
$\FoI$ with a
differential equation \labelcref{eq:re.dict.ode.X}, and relates $\FoI$ to
the generation interval distribution, $g(\aoi)$, via a convolution
[\cref{eq:re.dict.EL.i}].  If $g(\aoi)$ is not known, it is common to
assume it is a gamma distribution, as in (\protect\hypercref{eq:regamma}).
}


\ifepimomarxiv
  \clearpage
  \begingroup
  \makeatletter
  \@dblfptop=0pt
  \@fptop=0pt
  \makeatother
  \begin{table*}[!t]
  \footnotesize
  \caption{
    \tablecaptiontitle
    \hfill\break
    \tablecaptioncontent
  }\label{tab:models}
  \end{table*}
  \clearpage
  \endgroup
  \begin{table*}[!t]
  \fontsize{9}{10.8}\selectfont
\else
  \begin{table*}
  \footnotesize
  \caption{
    \tablecaptiontitle
    \hfill\break
    \tablecaptioncontent
  }\label{tab:models}
\fi

\newcounter{tableeq}
\renewcommand{\thetableeq}{T\arabic{tableeq}}
\crefname{tableeq}{equation}{equations}
\Crefname{tableeq}{Equation}{Equations}


\centerline{\normalsize\bfseries{SIR model}}

\modeltablerule
\medskip

\begin{multicols}{4}

\centerline{\bfseries Standard}
\mysqueeze

\makeatletter
\refstepcounter{tableeq}
\protected@edef\currentlabel{T\arabic{tableeq}}%
\label{eq:SIRraw}
\hypertarget{eq:SIRraw}{}
\makeatother

\begin{align}
  \ddt{S} &= - \tfrac{\beta}{N}\, S\,I \tag{\thetableeq a} \label{eq:SIR;S} \\
  \noalign{\vspace{8pt}}
  \ddt{I} &= \big(\tfrac{\beta}{N}\,S - \gamma\big) I \tag{\thetableeq b} \label{eq:SIR;I}
\end{align}

\columnbreak

\centerline{\bfseries\hspace{15pt}Dimensionless ($X\!=\!\tfrac{S}{N},\,Y\!=\!\tfrac{I}{N}$)}
\mysmallsqueeze

\makeatletter
\refstepcounter{tableeq}
\protected@edef\currentlabel{T\arabic{tableeq}}%
\label{eq:SIR}
\hypertarget{eq:SIR}{}
\makeatother
\begin{align}
  \ddtau{X} &= - \Rn\,X\,Y \tag{\thetableeq a} \label{eq:SIR;X} \\
  \noalign{\vspace{4pt}}
  \ddtau{Y} &= \big(\Rn\,X - 1\big) Y \tag{\thetableeq b} \label{eq:SIR;Y} \\
  \noalign{\vspace{4pt}}
  \inc &= \Rn\,X\,Y \tag{\thetableeq c} \label{eq:SIR;inc}
\end{align}

\columnbreak

\centerline{\bfseries Parameters}
\mysmallsqueeze

\makeatletter
\refstepcounter{tableeq}
\protected@edef\currentlabel{T\arabic{tableeq}}%
\label{eq:SIRparms}
\hypertarget{eq:SIRparms}{}
\makeatother

\begin{subequations}
\begin{align}
  \beta &= \scalebox{\modeltablescaleby}{\raisebox{2pt}{\parbox[c]{3cm}{\centering transmission rate}}}
           \tag{\thetableeq a} \label{eq:SIRparms;beta} \\
  \noalign{\vspace{2pt}}
  \tfrac{1}{\gamma} &= \scalebox{\modeltablescaleby}{\parbox[c]{3cm}{\centering mean infectious\\period}}
               \tag{\thetableeq b} \label{eq:SIRparms;gamma} \\
  \noalign{\vspace{2pt}}
  \gimean &= \scalebox{\modeltablescaleby}{\parbox[c]{3cm}{\centering mean generation\\interval}}
           = \tfrac{1}{\gamma} \tag{\thetableeq c} \label{eq:SIRparms;gimean} \\
  \noalign{\vspace{2pt}}
  \Rn &= \scalebox{\modeltablescaleby}{\parbox[c]{3cm}{\centering basic reproduction\\number}}
       = \tfrac{\beta}{\gamma} \tag{\thetableeq d} \label{eq:SIRparms;Rn} 
\end{align}
\end{subequations}

\columnbreak

\centerline{\bfseries Properties}
\mysqueeze

\null
\vfill

\makeatletter
\refstepcounter{tableeq}
\protected@edef\currentlabel{T\arabic{tableeq}}%
\label{eq:SIRproperties}
\hypertarget{eq:SIRproperties}{}
\makeatother
\begin{align}
  \tau &= \gamma\,t \tag{\thetableeq a} \label{eq:taudef} \\
  \noalign{\vspace{4pt}}
  g(\aoi) &= e^{-\aoi} \tag{\thetableeq b} \label{eq:giSIR} \\
  \noalign{\vspace{4pt}}
  \Lap{g}(\lambda) &= \tfrac{1}{\lambda+1} \tag{\thetableeq c} \label{eq:LapSIR} \\
  \noalign{\vspace{4pt}}
  \lampm &= \Rn\,\xpm - 1 \tag{\thetableeq d} \label{eq:lampmSIR}
\end{align}

\vfill
\null

\end{multicols}

\centerline{\normalsize\bfseries{SEIR model}}

\modeltablerule
\medskip

\begin{multicols}{4}

\centerline{\bfseries Standard}
\mysqueeze

\makeatletter
\refstepcounter{tableeq}
\protected@edef\currentlabel{T\arabic{tableeq}}%
\label{eq:SEIRraw}
\hypertarget{eq:SEIRraw}{}
\makeatother

\begin{align}
  \ddt{S} &= - \tfrac{\beta}{N}\, S\,I \tag{\thetableeq a} \label{eq:SEIR;S} \\
  \noalign{\vspace{8pt}}
  \ddt{E} &= \tfrac{\beta}{N}\,S\,I - \Erate E \tag{\thetableeq b} \label{eq:SEIR;E} \\
  \noalign{\vspace{8pt}}
  \ddt{I} &= \Erate E - \Irate I \tag{\thetableeq c} \label{eq:SEIR;I}
\end{align}

\columnbreak

\centerline{\bfseries\hspace{15pt}Dimensionless ($\Eprop\!=\!\tfrac{E}{N},\,\Iprop\!=\!\tfrac{I}{N}$)}
\mymedsqueeze
\vspace{10pt}

\makeatletter
\refstepcounter{tableeq}
\protected@edef\currentlabel{T\arabic{tableeq}}%
\label{eq:SEIR}
\hypertarget{eq:SEIR}{}
\makeatother

\begin{align}
  \ddtau{X} &= - \Rn\, X\,\Iprop \tag{\thetableeq a} \label{eq:SEIR;X} \\
  \noalign{\vspace{8pt}}
  \ddtau{\Eprop} &= \Rn\,X\,\Iprop - \frac{1}{\ell}\Eprop \tag{\thetableeq b} \label{eq:SEIR;Eprop} \\
  \noalign{\vspace{8pt}}
  \ddtau{\Iprop} &= \frac{1}{\ell}\Eprop - \Iprop \tag{\thetableeq c} \label{eq:SEIR;Iprop} \\
  \noalign{\vspace{8pt}}
  \inc &= \Rn\,X\,\Iprop \tag{\thetableeq d} \label{eq:SEIR;inc}
\end{align}

\columnbreak

\centerline{\bfseries Parameters}
\mysmallsqueeze

\makeatletter
\refstepcounter{tableeq}
\protected@edef\currentlabel{T\arabic{tableeq}}%
\label{eq:SEIRparms}
\hypertarget{eq:SEIRparms}{}
\makeatother

\begin{align*}
  \dfrac{1}{\Erate} &= \scalebox{\modeltablescaleby}{\parbox[c]{3cm}{\centering mean latent\\period}}
  \tag{\thetableeq a} \label{eq:SEIRparms;Erate}\\
  \noalign{\vspace{2pt}}
  \dfrac{1}{\Irate} &= \scalebox{\modeltablescaleby}{\parbox[c]{3cm}{\centering mean infectious\\period}}
  \tag{\thetableeq b} \label{eq:SEIRparms;Irate}\\
  \noalign{\vspace{2pt}}
  \gimean &= \dfrac{1}{\Erate} + \dfrac{1}{\Irate}
  \tag{\thetableeq c} \label{eq:SEIRparms;gimean}\\
  \noalign{\vspace{2pt}}
  \ell &= 
          \Irate/\Erate
  \tag{\thetableeq d} \label{eq:SEIRparms;ell}\\
  \noalign{\vspace{2pt}}
  \Rn &= \beta\,\Irate^{-1}
  \tag{\thetableeq e} \label{eq:SEIRparms;Rn}
\end{align*}

\columnbreak

\centerline{\bfseries Properties}
\mysqueeze

\null
\vfill

\makeatletter
\refstepcounter{tableeq}
\protected@edef\currentlabel{T\arabic{tableeq}}%
\label{eq:SEIRproperties}
\hypertarget{eq:SEIRproperties}{}
\makeatother

\begin{align}
  \tau &= \Irate\,t \tag{\thetableeq a} \label{eq:taudef.SEIR} \\
  \noalign{\vspace{6pt}}
  g(\aoi) &= 
    \begin{cases}
      \aoi e^{-\aoi}, & \scalebox{0.8}{\text{$\ell=1$}} \\
      \noalign{\vspace{3pt}}
      \frac{e^{-\aoi} - e^{-\aoi/\ell}}{1 - \ell}, & \scalebox{0.8}{\text{$\ell\ne1$}} 
    \end{cases}
    \tag{\thetableeq b} \label{eq:giSEIR} \\
  \noalign{\vspace{6pt}}
  \Lap{g}&(\lambda) = \tfrac{1}{\ell\lambda+1}\cdot\tfrac{1}{\lambda+1} \tag{\thetableeq c} \label{eq:LapSEIR} \\
  \noalign{\vspace{4pt}}
  &\hspace{-0.9cm}\lampm = \text{\scalebox{0.65}{
    $\dfrac{2(\Rn\xpm-1)}{\sqrt{(1-\ell)^2 + 4\ell\Rn\xpm} + (1+\ell)}$
  }}
  \tag{\thetableeq d} \label{eq:lampmSEIR}
\end{align}

\vfill
\null

\end{multicols}


\centerline{\normalsize\bfseries{Renewal equation}}

\modeltablerule
\medskip

\begin{multicols}{3}


\makeatletter
\refstepcounter{tableeq}
\protected@edef\currentlabel{T\arabic{tableeq}}%
\label{eq:re.dict}
\hypertarget{eq:re.dict}{}
\makeatother

\centerline{\bfseries Dimensionless renewal equation}

\begin{align}
  \dd{X}{\tau} &= -X(\tau)\FoI(\tau)
  \tag{\thetableeq a} \label{eq:re.dict.ode.X} \\
  \noalign{\vspace{4pt}}
  \FoI(\tau) &= \Rn \int_0^\infty X(\tau-\aoi)\FoI(\tau-\aoi)g(\aoi)\,\dee\aoi
  \tag{\thetableeq b} \label{eq:re.dict.EL.i} \\
  \noalign{\vspace{6pt}}
  \inc &= X\,\FoI
  \tag{\thetableeq c} \label{eq:re.dict;inc}
\end{align}


\columnbreak

\centerline{\bfseries For general $g(\aoi)$}

\makeatletter
\refstepcounter{tableeq}
\protected@edef\currentlabel{T\arabic{tableeq}}%
\label{eq:reproperties}
\hypertarget{eq:reproperties}{}
\makeatother
\begin{align}
  \gimean &= \int_0^\infty \aoi\,g(\aoi)\,\dee\aoi
  \tag{\thetableeq a} \label{eq:re.mean} \\
  \noalign{\vspace{8pt}}
  \tau &= t/\gimean
  \tag{\thetableeq b} \label{eq:re.tau} \\
  \noalign{\vspace{4pt}}
  \frac{1}{\Rn\xpm} &= \Lap{g}(\lampm)
  \tag{\thetableeq c} \label{eq:re.lamgeneric}
\end{align}

\columnbreak

\centerline{\bfseries For Gamma $g(\aoi)$ \  [$a=\frac{\gimean^2}{\gisd^2}$, $b=\frac{\gimean}{\gisd^2}$]}

\makeatletter
\refstepcounter{tableeq}
\protected@edef\currentlabel{T\arabic{tableeq}}%
\label{eq:regamma}
\hypertarget{eq:regamma}{}
\makeatother

\begin{align}
  g(\aoi) &=
  \frac{b^{a}}{\Gamma(a)}
  \aoi^{a-1}
             e^{-b\,\aoi}
  \tag{\thetableeq a} \label{eq:giRE.gamma} \\
  \noalign{\vspace{2pt}}
  \Lap{g}(\lambda) &=
  \left(\tfrac{b}{\lambda + b}\right)^{a} \tag{\thetableeq b} \label{eq:Lap.gamma} \\
  \noalign{\vspace{2pt}}
  \lampm &=
    b \big(\!(\Rn\xpm)^{1/a} - 1\!\big)
  \tag{\thetableeq c} \label{eq:lampmgamma}
\end{align}

\vfill
\null

\end{multicols}


\end{table*}
\ifepimomarxiv
  \clearpage
\fi



